\documentclass[12pt]{article}

\usepackage{amsmath,amssymb,amsfonts,theorem,epsfig,slashed,shadow,bbold,youngtab,yhmath,hyperref}
\usepackage{verbatim,rotate}

\newcommand{\bm}[1]{\mbox{\boldmath $ #1 $}}

\newcommand{\pa}{\partial}

\newcommand{\nn}{\nonumber}

\newcommand{\xo}{\stackrel{\circ}{x}\!{}}

\newcommand{\DI}{{\bf d}}
\newcommand{\DE}{{\bm \delta}}
\newcommand{\TR}{{\bf tr}}
\newcommand{\GI}{{\bf g}}
\newcommand{\wt}[1]{\widetilde{#1}}

\def\bea{\begin{eqnarray}}
\def\eea{\end{eqnarray}}
\def\be{\begin{equation}}
\def\ee{\end{equation}}
\def\ba{\begin{align}}
\def\ea{\end{align}}
\def\bse{\begin{subequations}}
\def\ese{\end{subequations}}
\def\ba{\left(\begin{array}{cc}}
\def\ea{\end{array}\right) }
\def\bv{\left(\begin{array}{c}}
\def\ev{\end{array}\right) }

\def\a{\alpha}
\def\b{\beta}

\def\g{\gamma}

\def\e{\epsilon}

\theoremstyle{plain}

\def\sideremark#1{\ifvmode\leavevmode\fi\vadjust{\vbox to0pt{\vss
 \hbox to 0pt{\hskip\hsize\hskip1em
 \vbox{\hsize3cm\tiny\raggedright\pretolerance10000
  \noindent #1\hfill}\hss}\vbox to8pt{\vfil}\vss}}}

\begin{document}
\thispagestyle{empty}

\vspace{.8cm}
\setcounter{footnote}{0}
\begin{center}
\vspace{-25mm}
{\Large
 {\bf  Quaternionic K\"ahler Detour Complexes \& \\[4mm] ${\bf {\cal N} =2}$ Supersymmetric Black Holes}\\[12mm]

 {\sc \small
     D.~Cherney$^{\mathfrak C}$, E.~Latini$^{\mathfrak L}$,  and A.~Waldron$^{\mathfrak W}$\\[4mm]

            {\em\small${}^{\mathfrak C,\mathfrak W}\!$
            Department of Mathematics\\ 
            University of California,
            Davis CA 95616, USA\\
            {\tt cherney,wally@math.ucdavis.edu}\\[2mm]
           ${}^\mathfrak{L}$ 
Department of Mathematics\\ 
            University of California,
            Davis CA 95616, USA\\
and\\ INFN, Laboratori Nazionali di Frascati, CP 13, I-00044 Frascati, Italy \\
{\tt emanuele@math.ucdavis.edu, latini@lnf.infn.it}\\[2mm]

            }}

 }

\bigskip

{\sc Abstract}\\[-4mm]
\end{center}

{\small
\begin{quote}

We study a class of supersymmetric spinning particle models derived from the radial quantization of stationary, spherically symmetric black holes of four dimensional ${\cal N} = 2$ supergravities. By virtue of the $c$-map, these spinning particles move in quaternionic K\"ahler manifolds. Their spinning degrees of freedom describe  mini-superspace-reduced supergravity fermions. We quantize these models using BRST detour complex technology. The construction of a nilpotent BRST charge is achieved by using local (worldline) supersymmetry ghosts to generating special holonomy transformations. (An interesting byproduct of the construction is a novel Dirac operator on the superghost extended Hilbert space.)
The resulting quantized models are gauge invariant field theories with fields equaling sections of special quaternionic vector bundles. They underly and generalize the quaternionic version of Dolbeault cohomology discovered by Baston. In fact, Baston's complex is related to the BPS sector of the models we write down. Our results rely on a calculus of operators on quaternionic K\"ahler manifolds that follows from BRST machinery, and although directly motivated by black hole physics, can be broadly applied to any model relying on quaternionic geometry.

\end{quote}
}

\newpage

\tableofcontents

\section{Introduction}

The main result of this paper is a detour complex for quaternionic K\"ahler manifolds. In physics language, 
this amounts to a gauge theory of 
higher (quaternionic) ``forms'' 
on these manifolds. 
To be precise,
we utilize special holonomy to split the tangent bundle of a 
$4n$-dimensional quaternionic K\"ahler manifold~$M$ into a product of
rank~2 and~$2n$ vector bundles~$H$ and $E$~\cite{Salamon},
$$
TM\cong E\otimes H\, ,
$$ 
and present an equation of motion and gauge invariances for
sections of 
$\wedge E$ (or, more generally, $\wedge E \otimes \odot H$).  

The results of the paper will appeal to multiple audiences including:
(i) Those readers interested in the differential geometry of quaternionic K\"ahler spaces. 
(ii) Readers studying various
supersymmetric quantum mechanical and spinning particle models in quaternionic K\"ahler and hyperK\"ahler backgrounds (such as such as gravitational instanton moduli spaces~\cite{ADHM}, Hitchin's moduli space of stable Higgs bundles~\cite{Hitchin}, geometric Langlands theory~\cite{Kapustin:2006pk} and hypermultiplet moduli spaces~\cite{Bagger:1983tt}, to name a few). 
(iii) Readers looking for applications of 
the BRST detour quantization of orthosymplectic constraint algebras developed for applications to higher spin systems in~\cite{Cherney:2009md}, on which these results heavily rely.
(iv)
Readers  wanting to apply our results to supergravity (SUGRA) black hole quantization since, remarkably, 
the mathematical structure presented above is exactly what is called for when studying the minisuperspace quantization of ${\cal N}=2$ SUGRA black holes~\cite{Gunaydin:2007bg,Pioline:2006ni}. 
(In particular, wavefunctions valued in~$\wedge E$ describe the fermionic degrees of freedom of these models.)
Therefore the paper is structured so that any of these readerships can easily extract the information they need.

In section~\ref{detour}, we introduce the notion of a detour complex, 
beginning with simple examples.
We then 
generalize our previous results on 
K\"ahler detour complexes to hyperK\"ahler manifolds.  
This result follows immediately from 
an isomorphism between
super Lie algebras of geometric operators 
mapping
Dolbeault and Lefschetz operators
on  
 K\"ahler forms
to their hyperK\"ahler analogues 
acting on sections of $\wedge E$. 
We then explain a main difficulty  solved in this paper: 
the construction of a geometric detour complex for quaternionic K\"ahler manifolds is seemingly obstructed by the higher rank of the analogous geometric super algebra. 
This problem is overcome in later sections by understanding the 
key {\it r\^ole} played by the BRST superghosts in the description of  quaternionic geometry. 
The main requisite geometric data is presented in section~\ref{special} together with our notations and conventions.

In Section~\ref{BH} we 
review
the relationship between quaternionic K\"ahler spinning particles and four dimensional black holes; the original motivation for this work. The latter can be described by  a spinning particle model coming from the minisuperspace reduction of 
${\cal N}=2$ supergravities~\cite{Gunaydin:2007bg}. 
The ``BPS'' conditions of this spinning particle model ({\it i.e.}, requiring solutions for which the local fermion supersymmetry transformations
vanish)  equal the reduction of the analogous conditions in the four dimensional SUGRA. Since those conditions amount to the attractor mechanism~\cite{Ferrara:1996um}
for four dimensional supersymmetric black holes, the quantized spinning particle model is an excellent laboratory for studying these objects\footnote{A very useful introduction to BPS black holes and the attractor mechanism is~\cite{Pioline:2006ni} (the formulation in~\cite{Denef:2000nb} also fits our viewpoint well).}. In particular,  it allows a  minisuperspace analysis of the Ooguri--Strominger--Vafa conjecture~\cite{Ooguri:2004zv} and the relationship between black hole wave functions and vacuum selection in string theory~\cite{Ooguri:2005vr}. This equivalence between the attractor flow equation and supersymmetric geodesic motion was observed in~\cite{Gutperle:2000ve,Gunaydin:2007bg}.

%

The introduction of BRST techniques to solve what could be stated as a purely geometrical problem suggests the presence of an
underlying gauge invariant physical model. 
This is indeed the case. The first of the relevant models is a hyperK\"ahler supersymmetric quantum mechanics.
This model can be enhanced to include quaternionic K\"ahler backgrounds once its four worldline supersymmetries are gauged. This yields
a supersymmetric spinning particle model consistent in any quaternionic K\"ahler manifold. We describe these models in sections~\ref{sigma} and~\ref{QKSUGRA}, respectively. 

Sections~\ref{special},~\ref{calculus} and~\ref{QKdetour} 
can in principle be read by geometers in isolation
from the other more physical sections. 
In section~\ref{calculus}, we give a calculus of geometric operators acting on sections of~$\wedge E$. Although, we were motivated to 
write these operators for quantum mechanical BRST reasons, the results themselves are purely geometric. They form the basic building
blocks of our quaternionic detour complex. They also place in a much more general setting the Dirac, Dirac--Fueter and detour operator
employed some time ago by Baston~\cite{MR1165872}. 

Finally our main result is given in section~\ref{QKdetour}, orchestrating all the previous results to build a gauge invariant,
higher ``form'' quantum field theory on quaternionic K\"ahler manifolds. 
It relies on the construction of a nilpotent BRST charge given in section~\ref{BRST} achieved by
utilizing the supersymmetry ghosts to 
generate special holonomy transformations. An interesting byproduct of this computation is a novel
Dirac operator on the BRST superghost Hilbert space. 

Asides from providing an explicit quantization of
the fermion modes of minisuperspace ${\cal N}=2 $ supersymmetric black holes, our quaternionic detour complex  has many potential further applications
and generalizations. In particular, it is closely related to the twistor methods of~\cite{Neitzke:2007ke}. Also, in some sense, the model is a higher spin
theory, so the methods of Vasiliev may be applicable to writing interactions for infinite towers of these quantum fields (see~\cite{Bekaert:2005vh}
 for an excellent review of these methods). Given
the existence of the underlying SUGRA theory, this is a very tantalizing possiblity. These and other directions for future work are 
discussed in the conclusions.


\section{Detour Complexes}

\label{detour}
The simplest example of a geometrical
detour complex  is given by the superalgebra, on any Riemannian manifold $M$, generated by the exterior derivative~${\bm d}$ and the codifferential ${\bm \delta}$:
\be
\{{\bm \delta},{\bm d}\}={\bm {\bm \Delta}}\, .~\label{ddd}
\ee
Here, the right hand side is the  form Laplacian which is a central element of this algebra. These operators act on differential forms $\Psi \in \Gamma(\wedge M)$, which may be viewed as
wavefunctions of an ${\cal N}=2$ supersymmetric quantum mechanical model~\cite{Witten:1982im}, 
with ${\bm {\bm \Delta}}$  the Hamiltonian and $({\bm \delta, \bm d})$ the two supercharges. Gauging the corresponding worldline translation and supersymmetries
yields a spinning particle (or 1-dimensional SUGRA) model which can be quantized using BRST machinery. 
In mathematical terms this amounts to computing the Lie algebra cohomology of the superalgebra~\eqref{ddd}.

However, when defining Lie algebra cohomology for superalgebras, some care is needed~\cite{Fuchs}. In physics terms
this amounts to choices of vacua/polariz\-ations for commuting superghosts~\cite{Fuster:2005eg,Siegel:1990zf}. It turns out that a distinguished choice exists such that 
the cohomology is neatly arranged in terms of gauge invariances, Bianchi identities and the equations of motion of a  gauge invariant
field theory. In a higher spin setting this was first observed in the context of an unfolded formulation and what is called the ``twisted adjoint 
representation''~\cite{Twist}. 
(Very recently the unfolding technique has been shown to be equivalent to the BRST one~\cite{Gelfond:2010xs}. The idea of studying worldline descriptions of higher spin systems, via detour and path integral quantization has also been  analyzed in~\cite{Bastianelli:2009eh} and~\cite{Bastianelli:2008nm}.) In~\cite{Cherney:2009mf}
 we used a split 
choice of ghost polarization\footnote{The technique of split ghost polarizations is equivalent to the twisted adjoint representation of~\cite{Twist}. It has also been employed in~\cite{Barnich:2004cr,Campoleoni:2008jq}.} to construct detour complexes from constraint algebras. 
(For systems with anti-commuting ghosts, this method reproduces known results~\cite{Sorokin:2004ie} for totally symmetric higher spin fields).  
The term detour complex was chosen because the result of the BRST technology produced complexes of the type studied recently by conformal geometers.
The main idea being to connect standard complexes and their duals by (typically higher order in derivatives) detour operators~\cite{Branson:2003an,GHW}.
For the simplest case of the de Rham complex, the detour machinery yields a cohomology neatly encapsulated by the complex
\be\nn
\begin{array}{c}
\cdots
\stackrel{\bm d} {\longrightarrow}
\Lambda M
\stackrel{\bm d} {\longrightarrow}
\Lambda M
\stackrel{\bm d} {\longrightarrow}
\Lambda M
\rightarrow\cdots
\quad
\cdots\rightarrow 
\Lambda M
\stackrel{\bm \delta}{\longrightarrow} 
\Lambda M
\stackrel{\bm \delta} {\longrightarrow}
\Lambda M
\stackrel{\bm \delta} {\longrightarrow}
\cdots
\\
\ \Big|\hspace{-.8mm}\raisebox{-2.5mm}{\underline{\quad\quad\quad \ \raisebox{1mm}{\bm \delta \bm d}\quad\qquad\quad }} \hspace{-1.2mm}{\Big\uparrow}
\end{array}
\ee
The self-adjoint detour operator $\bm \delta \bm d$ encodes the equations of motion $\bm \delta \bm d  A=0$ of a $p$-form gauge field $A$ and connects the standard de Rham complex to its dual. These incoming and outgoing complexes encode the gauge and gauge for gauge symmetries, and Bianchi as well as Bianchi for Bianchi identities
of $p$-form electromagnetism.

A more sophisticated example is that of the K\"ahler detour complex; 
on these manifolds the exterior derivative and codifferential decompose into Dolbeault operators and their duals~\cite{Griff,FigueroaO'Farrill:1997ks}
$$
\bm d =
 \bm \partial + \bar {\bm \partial} \, ,
\qquad \bm \delta = \bm \partial^* 
+ \bar{ \bm \partial}^*\, ,
$$
subject to the superalgebra
$$
\{\bm\partial,\bm\partial^*\}=\frac12\bm {\bm \Delta}=\{\bar{\bm\partial},\bar{\bm\partial}^*\}\, .
$$
In addition, an  $\frak{sl}(2)$ Lefschetz algebra acts on the Dolbeault cohomology of a K\"ahler manifold~$M$. This corresponds to the $R$ symmetry algebra
of the above ${\cal N}=4$  superalgebra
$$
\left[\bm \Lambda,\begin{pmatrix}\bm \partial\\ \bar{\bm \partial}\end{pmatrix}\right]=\begin{pmatrix}\bar{\bm \partial}^*\\ -{\bm \partial}^*\end{pmatrix}\, ,\qquad
\left[\begin{pmatrix}\bar{\bm \partial}^*\\ -{\bm \partial}^*\end{pmatrix},\bm L\right]=\begin{pmatrix}\bm \partial\\ \bar{\bm \partial}\end{pmatrix}\, ,
$$
$$
[\bm H,\bm \Lambda] = -2\bm \Lambda\, ,\qquad	[\bm H,\bm L] = 2\bm L\, ,\qquad	[\bm \Lambda,\bm L] = \bm H\, .
$$
Differential forms on a K\"ahler manifold are bigraded by their holomorphic and antiholomorphic degrees $(p,q)$
in terms of which the eigenvalues of the operator  $\bm H$ are 
$p+q-{\frac12\,  \rm dim}\,M$. The operator $\bm \Lambda$ maps $(p,q)$ to
$(p-1,q-1)$-forms by contracting with the K\"ahler form and the operator $\bm L$ is its dual. 
 
The K\"ahler analog of $p$-form electromagnetism~\cite{Cherney:2009vg}
follows by a detour complex treatment of the spinning particle\footnote{ 
Supersymmetric mechanics on K\"ahler manifolds have been extensively studied in~\cite{Marcus,Fiorenzo} and~\cite{Bellucci:2001ax}.} model obtained by gauging worldline translations, supersymmetries and the $R$-symmetry $\bm \Lambda$. 
Nilpotentcy of 
$Q=\bm\partial \frac{\pa}{\pa p}+   \bar{\bm \partial}  \frac{\pa}{\pa \bar{p}}$
acting on polynomials in Grassmann even variables $p, \bar{p}$ with coefficients in 
$\wedge M$ yields the left hand side of the complex
$$
\scalebox{.8}{$
\scalebox{1.5}{$\!\!\cdots$}\quad 
\begin{array}{ccccccccccccccc}
\overset{\!\scalebox{.8}{ \bm \partial}}{ \searrow}&&&&&&&&&&&&\underset{\!\scalebox{.8}{$\bar{\bm \partial}^*$}}{ \nearrow}\\
                &\, \, \Lambda M \, \, &&&&&&&&&&\, \, \Lambda M \, \, \\
\underset{\!\scalebox{.8}{$\bar{\bm \partial}$}}{ \nearrow}&&\overset{\!\scalebox{.8}{ \bm \partial}}{ \searrow}&&&&&&&&\underset{\!\scalebox{.8}{$\bar{\bm \partial}^*$}}{ \nearrow}&&\overset{\!\scalebox{.8}{ \bm \partial}^*}{ \searrow}\\
&&&\, \, \Lambda M \, \,&& &&&&\, \, \Lambda M \, \,&&\\
\overset{\!\scalebox{.8}{ \bm \partial}}{ \searrow}&&\underset{\!\scalebox{.8}{$\bar{\bm \partial}$}}{ \nearrow}&&\overset{\!\scalebox{.8}{ \bm \partial}}{ \searrow}&&&&\underset{\!\scalebox{.8}{$\bar{\bm \partial}^*$}}{ \nearrow}&&\overset{\!\scalebox{.8}{ \bm \partial}^*}{ \searrow}&&\underset{\!\scalebox{.8}{$\bar{\bm \partial}^*$}}{ \nearrow}\\
                &\, \, \Lambda M \, \, &&&& \, \, \Lambda M \, \,&\overset{\!\scalebox{.9}{\bm G}}{-\!-\!\!\!\longrightarrow}&\, \, \Lambda M \, \, &&&&\, \, \Lambda M \, \, 
                 \\
\underset{\!\scalebox{.8}{$\bar{\bm \partial}$}}{ \nearrow}&&\overset{\!\scalebox{.8}{ \bm \partial}}{ \searrow}&&\underset{\!\scalebox{.8}{$\bar{\bm \partial}$}}{ \nearrow}&&&&\overset{\!\scalebox{.8}{ \bm \partial}^*}{ \searrow}&&\underset{\!\scalebox{.8}{$\bar{\bm \partial}^*$}}{ \nearrow}&&\overset{\!\scalebox{.8}{ \bm \partial}^*}{ \searrow}\\
&&&\, \, \Lambda M \, \, &&&&&&\, \, \Lambda M \, \,&& \\
\overset{\!\scalebox{.8}{ \bm \partial}}{ \searrow}&&\underset{\!\scalebox{.8}{$\bar{\bm \partial}$}}{ \nearrow}&&&&&&&&\overset{\!\scalebox{.8}{ \bm \partial}^*}{ \searrow}&&\underset{\!\scalebox{.8}{$\bar{\bm \partial}^*$}}{ \nearrow}\\
                &\, \, \Lambda M \, \, &&&&&&&&&&\, \, \Lambda M \, \, \\
\underset{\!\scalebox{.8}{$\bar{\bm \partial}$}}{ \nearrow}&&&&&&&&&&&&\overset{\!\scalebox{.8}{ \bm \partial}^*}{ \searrow}
\end{array}\quad\scalebox{1.5}{$\cdots$}
$}
$$
Upon fixing a dimension for $M$ and a bi-grading $(p,q)$ 
 this incoming complex becomes the Hodge diamond from complex manifold 
theory. It may be interpreted as gauge (and gauge for gauge) invariances of the ``long'' or detour  operator $\bm G$. 
Explicitly, gauge invariance reads
$$
A\to A+\bm \partial \alpha +\bar{\bm \partial} \bar \alpha\, .
$$
Clearly the equations $\bm \partial \bar{\bm \partial} A=0$ are invariant, yet potentially over or underdetermined. 
Taking the K\"ahler trace yields the desired 
equations of motion $\bm\Lambda \bm \partial \bar{\bm \partial} A=0$. However, the operator 
$\bm \Lambda \bm \partial \bar{\bm \partial}$ is not self-adjoint
and so does not naturally connect the ``incoming'' Dolbeault complex with the ``outgoing''  dual complex depicted on the right hand side above.  
The self adjoint operator 
$$
\bm G=\  \scalebox{1.2}{:} I_{0}(2\sqrt{\bm L\bm \Lambda})\, (\bm {\bm \Delta} - 2\bm\partial\bm\partial^*-2\bar{\bm \partial}\bar{\bm \partial}^*)
+\ 2\, \frac{I_{1}(2\sqrt{\bm L\bm\Lambda})}{\sqrt{\bm L\bm \Lambda}} \, (\bm\partial\bar{\bm \partial}\, \bm\Lambda +\bm L\, \bm \partial^*\bar{\bm \partial}^* )
\, \scalebox{1.2}{:}\, \ \ 
$$
found in~\cite{Cherney:2009vg} gives an equivalent equation of motion $\bm G A=0$.
Here $:\bullet:$ denotes normal ordering of $\bullet$ by form degree 
and the functional dependence on $\bm L\bm\Lambda$ through the
modified Bessel functions of the second kind is analytic at the origin. 


In the special case that $M$ is hyperK\"ahler, replacing differential forms by sections of 
$\wedge E$ gives another representation of the above  ${\cal N}=4$
supersymmetry algebra:
The tangent bundle $TM$ for $4n$-dimensional manifolds $M$ with quaternionic holonomy  splits into a product of vector bundles
$$
TM\cong H\otimes E\, ,
$$
of rank $2$ and $2n$, respectively. 
The connection on a  hyperK\"ahler manifold 
acts  on sections  $X^\alpha$ and $X^A$ of $H$ and $E$, respectively,  as
\be\nn
\nabla X^\alpha=d X^\alpha  +  \omega^\a_\b X^\b \  ,\qquad
\nabla X^A=d X^A+\Omega^A_B X^B\, ,
\ee
where the  one-form $\Omega^A_B$ is $\frak{sp}(2n)$-valued. 
Writing the Levi-Civita connection as $\nabla^{\alpha A}$ in a basis for $H\otimes E$,
there are $\mathfrak{sp}(2)$ doublets of
 exterior derivatives and codifferentials acting on $\wedge E$ via
\bea\nn
{\bf d}^\alpha: & X^{A_1\ldots A_k}&\mapsto \nabla^{\alpha [A_1} X^{A_2\ldots A_{k+1}]}\, ,\\\nn
{\bm \delta}_{\alpha} : & X^{A_1\ldots A_k}&\mapsto 
k \nabla_{\alpha A} X^{A A_1\ldots A_{k-1}}\ \, ,
\eea
in the index notation explained in Section~\ref{special}. They obey the ${\cal N}=4$ algebra
\bea\nn
\{\bf d^\alpha, \bf d^\beta\}=&0&=\{\bm \delta_\alpha, \bm \delta_\beta\}\, ,\\[3mm]\nn
\{\bm \delta_\alpha, \bm d^\beta\}&=& - \frac 12 \delta^\beta_\alpha  {\bm \Delta}\, ,
\eea
where ${\bm \Delta}$ is the Bochner Laplacian $\nabla_\mu \nabla^\mu$. 
Only an $\frak{sp}(2)$ subalgebra of the $\frak{so}(2,2)$ $R$-symmetry of this 
${\cal N}=4$ 
superalgebra acts non-trivially in this hyperK\"ahler representation. 
The non-trivial $R$-symmetries are built from
the $\frak{sp}(2n)$ invariant tensor $J$
\bea\nn
&{\bf g}: X^{A_1\ldots A_k}&\mapsto J^{[A_1A_2} X^{A_3\ldots A_{k+2}]}\, ,\\[2mm]\nn
&{\bf N}: X^{A_1\ldots A_k}&\mapsto  k\, X^{A_1\ldots A_k}\, ,\\[2mm]\nn
&{\bf tr}: X^{A_1\ldots A_k}&\mapsto k(k-1)\, J_{AB} X^{BAA_1\ldots A_{k-2}}\, ,
\eea
and obey the algebra
$$[{\bf tr},{\bf N}]=2\, {\bf tr}\, ,\qquad [{\bf tr},{\bf g}]=4({\bf N} -n)\, ,\qquad [{\bf N},{\bf g}]=2\, {\bf g}\, ,$$
$$\ [{\bm \delta}_\alpha,{\bf N}]\  =\ {\bm \delta}_\alpha\, , \qquad \  [{\bf N},{\bf d}^\alpha]\ =\ {\bf d}^\alpha\, ,$$
$$[{\bf tr},{\bf d}_\alpha]=2\, {\bm \delta}_\alpha\, ,\qquad[{\bm \delta}^\alpha,{\bf g}]=2\, {\bf d}^\alpha\, .$$

The dictionary 
$$
{\bm d}^\alpha\leftrightarrow\begin{pmatrix}\bm \partial\\  \bar{\bm \partial}\end{pmatrix}\, ,\quad
{\bm \delta}_\alpha\leftrightarrow\begin{pmatrix} - \bm \partial^* & -\bar{\bm \partial^*}\end{pmatrix}\, ,\qquad
\bm g \leftrightarrow 2\bm L \, ,\quad  {\bf tr} \leftrightarrow 2\bm \Lambda\, ,
$$
between the K\"ahler and hyperK\"ahler representations of the ${\cal N}=4$ superalgebra 
allows the K\"ahler detour complex to be translated directly to a hyperK\"ahler one. 

In particular, nilpotence of the operator
$Q={\bm d}^\alpha\frac{\partial}{\partial p^\alpha}$ on polynomials in the Grassmann even variables $p^\alpha$ with coefficients in $\Gamma(\wedge E)$ gives gauge and gauge for gauge invariances of 
the over-determined,
Maxwell like, and Einstein versions of the hyperK\"ahler equations of motion
\bea \nn
\bm d_\alpha \bm d^\alpha A = 0\, \Rightarrow ~{\bf tr}\,  \bm d_\alpha \bm d^\alpha A =0 \, \Leftrightarrow \bm G A=0 \, ,~~~~~~~~~~~~~~~~~~~~ \\[3mm] \nn
\vspace{5.5cm}
\bm G=\  \scalebox{1.2}{:} I_{0}(\sqrt{\bf g\, tr})\, 
(\bm {\bm \Delta} +  2 \, {\bf d}_\alpha {\bm \delta}^\alpha)
-\ 2\, \frac{I_{1}(\sqrt{\bf g\, tr})}{\sqrt{\bf g\, tr}} \, 
({\bm d}_\alpha {\bm d}^\alpha\, {\bf tr} +\bm g\, {\bm \delta}_\alpha {\bm \delta}^\alpha )
\, \scalebox{1.2}{:}\, \ \ ,
\eea
for gauge fields $A\in \Gamma(\wedge E)$. 
Explicitly, the gauge invariance reads
$$
A\to A+ \bm d ^\alpha \alpha_\alpha\, .
$$
The equation of motion $\bm d_\alpha \bm d^\alpha A = 0$ was first generalized to the more complicated quaternionic K\"ahler case by Baston~\cite{MR1165872}, and later recovered in the context of BPS, 
${\cal N} = 2$ supersymmetric black hole systems in~\cite{Neitzke:2007ke}. 
The main result of this paper is to further extend this generalization to the full ``Einstein'' equations of motion 
$\bm G A=0$ in the quaternionic K\"ahler setting. It relies on a trio of geometric operators (one of which is 
Baston's original second order operator) transforming as a triplet under 
$\frak{sp}(2)$ $R$-symmetries.
We now present the
basic geometric data on quaternionic K\"ahler manifolds needed for this paper.




\section{Special Geometry}


\label{special}

HyperK\"ahler and quaternionic K\"ahler manifolds in dimension $4n$ and signature
$(2n,2n)$ enjoy $\mathfrak{sp}(2n)$ and $\mathfrak{sp}(2)\otimes \mathfrak{sp}(2n)$ 
holonomy, respectively.\footnote{The maximally split signature corresponds to
paraquaternionic holonomy -- all our results apply to general signatures, this choice
being a matter of notational convenience.}
In either case, this implies that the tangent bundle 
splits into a product of vector bundles~\cite{Bagger:1983tt}

$$
TM\cong H\otimes E
$$
of rank $2$ and $2n$, respectively. Therefore, we denote 
curved and flat indices by $\mu,\nu,\ldots$ and $m,n,\ldots$ respectively, and
decompose
tangent space indices 
as
\be\nn
m=\alpha  A \, ,
\ee
where $A=1,\ldots, 2n$ and $\alpha=1,2$ label the fundamental representations
of $\mathfrak{sp}(2n)$ and~$\mathfrak{sp}(2)$, respectively. 

The invariant $\mathfrak{so}(2n,2n)$ metric decomposes this way as
\be\nn
\eta_{mn}=\varepsilon_{\alpha\beta} J_{AB}\, ,
\ee
where $\varepsilon_{\alpha\beta}$ and $J_{AB}$ are the 
$\mathfrak{sp}(2)$ and $\mathfrak{sp}(2n)$ invariant,
antisymmetric tensors. 
This allows for all indices to be raised and lowered independently.
For example, 
$v_A\equiv J_{AB}v^B$,  $v^\alpha\equiv v_\beta \varepsilon^{\beta\alpha}$
and  $\varepsilon^\alpha{}_\beta = \delta^\alpha_\beta=-\varepsilon_\beta{}^\alpha$. Note that we use an uphill convention.


%
The action of the connection on  sections of $H$ and $E$, respectively, is given~by
\be\nn
\nabla X^\alpha=d X^\alpha+\omega^\alpha_\beta X^\beta\, ,\qquad
\nabla X^A=d X^A+\Omega^A_B X^B\, ,\\
\ee
where both $\omega_{\a\b}$ and $\Omega_{AB}$ are symmetric. 
On hyperK\"ahler manifolds, only the latter is non-zero. 
This may be extended to arbitrary tensor products of sections of $H$ and $E$ in the obvious way. For the purposes of calculations involving such products, we specify this action by
introducing representations of 
the $\frak{sp}(2n)$ and $\frak{sp}(2)$ subalgebras of the full local Lorentz 
algebra $\frak{so}(2n,2n)$.  
The generators of these algebras are represented as operators $T^{AB}$ and $t^{\a\b}$, indexed by symmetric pairs of indices, that act on  
$\mathfrak{sp}(2n)$ and $\mathfrak{sp}(2)$ indices by 
\bea \nn
T^{AB} X^{C} &=&  J^{CA} X^{B }+J^{CB} X^A\, ,\\[2mm] 
t^{\a\b} X^{\g }\ &=&\ \e^{\g\a} X^{ \b }+\e^{\g\b} X^{ \a }\, .
\label{Tt}
\eea
These operators satisfy
\be\nn
[T^{AB},T^{CD}]=J^{CA}T^{BD}+J^{CB}T^{AD}+J^{DA}T^{BC}+J^{DB}T^{AC}\, ,
\ee
\be\nn
[t^{\alpha\beta},t^{\gamma\delta}]=\varepsilon^{\gamma\alpha}t^{\beta\delta}+\varepsilon^{\gamma\beta}t^{\alpha\delta}+\varepsilon^{\delta\alpha}t^{\beta\gamma}+\varepsilon^{\delta\beta}t^{\alpha\gamma}\, ,
\ee
their extension to higher tensors is by the usual Leibnitz rule, and thus
\bea\nn
\nabla =d +\frac12 \, \omega^{\alpha}_{\beta} \, t_{\a}^{\b}  + \frac12\,  \Omega^{A}_{B}\,  T_{A}^{B} \, .
\eea
Throughout this paper, the symbol $\nabla$ will refer to this definition.

The final geometric ingredient needed here is the Riemann tensor.
As a result of special holonomy it has the decomposition~\cite{Bagger:1983tt}
\be \label{Riemann}
R_{\a A \, \b B \, \g C \, \delta D }=\Lambda \varepsilon_{(\alpha|\gamma|}\varepsilon_{\beta)\delta} J_{AB}J_{CD}
                 + \varepsilon_{\alpha\beta}\varepsilon_{\gamma\delta}
                   [\Lambda J_{(A|C|}J_{B)D}
                 +\Omega_{ABCD}]\, .
\ee
Hence,
 the commutator of covariant derivatives 
 on sections of $H$ and $E$  follows from:
\be\nn
[\nabla_{A\alpha},\nabla_{B\beta}]\, \phi_{C\gamma}
=\Lambda J_{BA}\,  \varepsilon_{\gamma(\alpha}\phi_{C\beta)}
+\Lambda \varepsilon_{\beta\alpha}\,  J_{C(A}\phi_{B)\gamma}
+\varepsilon_{\beta\alpha}\, \Omega^D_{ABC}\phi_{D\gamma}
\, .
\ee
This specifies an action on higher rank tensors which can be succinctly expressed in terms of the operators
\be\nn
[\nabla_{A\alpha},\nabla_{B\beta}]
=\frac12 \,J_{BA}\, t_{\alpha\beta}\,  + \frac12 \, \varepsilon_{\beta \alpha}\, \Big(T_{AB} 
+  
\Omega_{ABC}^D T^{C}_D\Big)\, .
\ee
The tensor $\Omega_{ABCD}$ is totally symmetric and will appear only seldomly in 
this paper since it cannot couple to the antisymmetric sections of $\wedge E$ which appear in our models.
The terms proportional to the constant $\Lambda$ are present only on quaternionic K\"ahler
manifolds and vanish for the hyperK\"ahler case\footnote{Note that
these are not proportional to $\eta_{r[m}\eta_{n]s}$ -- the constant curvature Riemann
tensor -- since general quaternionic K\"ahler manifolds are not constant curvature.}. Finally, note that the Ricci and scalar curvatures
are
$
R_{mn}=-\Lambda (n+2)\eta_{mn}$ and
$R=-4\Lambda n(n+2)$.


\section{${\cal N}=2$ Supersymmetric Black Holes and Quaternionic Geometry}

\label{BH}
Breitenlohner, Maison and Gibbons~\cite{Breitenlohner:1987dg} showed that Kaluza--Klein reduction along a single isometry of 
a four dimensional, curved space non-linear sigma models coupled to Maxwell fields
$$
S=-\frac12 \int \Big[d^4x \sqrt{-\rm g} \, R +  g^{(4)}_{\cal AB}(\varphi) d \varphi^{\cal A}\wedge{}^*d\varphi^{\cal B} + \frac12 F^{\cal I}\wedge
\Big(M_{\cal IJ} \, {}^* F^{\cal J} + N_{\cal IJ} F^{\cal J}\Big) \Big]\, ,
$$
(where ${\cal A,B}=1,\ldots,n_S$ the number of scalar fields and ${\cal I,J}=1,\ldots, n_V$ the number of vector fields)
yields a three dimensional curved space non-linear sigma model
$$
S=-\frac12 \int \Big[d^3x \sqrt{-\rm g} \, R +  g_{\mu\nu}(\phi) d \phi^{\mu}\wedge{}^*d\varphi^{\nu}\Big]\, .
$$
The metric $g_{\mu\nu}$ on the moduli space of the three dimensional non-linear sigma model 
depends on  that of the four dimensional sigma model $g^{(4)}_{\cal AB}$ as well as the  couplings $M_{\cal IJ}$ and $N_{\cal IJ}$ of the Maxwell field strengths $F^{\cal I}$ to the four dimensional scalars $\varphi^{\cal A}$. We refer to the original paper~\cite{Breitenlohner:1987dg} for the precise formul\ae. Suffice it to say, that
the $n_S$ scalars in four dimensions are enlarged to a set of $n_S+2n_V+2$ scalars coming from the dilaton, dualized graviphoton, Maxwell Kaluza--Klein
scalar  modes and dualized three Maxwell fields. They span the moduli space ${\cal M}$ of the three dimensional sigma model, and in this paper 
we will be primarily interested in the case that $\dim {\cal M}=4n$. In particular when the original four dimensional theory is the bosonic sector
of ${\cal N}=2$ SUGRA, the four dimensional scalar moduli space is a K\"ahler manifold and its image under dimensional reduction 
is a (para)quaternionic K\"ahler manifold. This correspondence is known as the $c$-map~\cite{Ferrara:1989ik,Gunaydin:1983rk,Gunaydin:1983bi,Cecotti:1988qn}.

When the reduction isometry is generated by a timelike Killing vector, solutions of the three dimensional sigma model correspond to stationary
solutions of the four dimensional theory. If we make the additional assumption of spherical symmetry of  the three dimensional stationary slices
$$
ds^2 = N^2(\rho)d\rho^2 + r^2(\rho) (d\theta^2 + \sin^2\theta\,  d\varphi^2)\, ,
$$ 
solutions then derive from a one dimensional action 
$$
S=-\frac{1}{2}\int d\rho \Big[N+N^{-1}(r'^2-r^2\phi'^\mu g_{\mu\nu} \phi'^\nu)\Big]\,  ,
$$
where primes denote $\rho$-derivatives. This model can be interpreted as a relativistic particle moving in a cone metric
$$
dr^2 - r^2 d\phi^\mu g_{\mu\nu} d\phi^\nu \, ,
$$
over the quaternionic K\"ahler moduli space ${\cal M}$. Classical solutions separate into radial motion and geodesics on the moduli space ${\cal M}$.
Of these, the extremal black hole solutions of the original four-dimensional theory are necessarily in correspondence with lightlike geodesics~\cite{Breitenlohner:1987dg}; 
the radial quantization of static, spherically symmetric black holes in Einstein and 
Einstein-Maxwell gravity has been studied in~\cite{Thiemann:1992jj}. 
The consequences of the four dimensional local supersymmetry of the underlying ${\cal N}=2$ SUGRA
can be incorporated in this minisuperspace approximation by computing the dimensional reduction of the supersymmetry
transformations (see~\cite{Gunaydin:2007bg}). BPS states follow by requiring that the transformations of the fermions vanish. This requirement splits into a radial condition
$$
dr=Nd\rho\, ,
$$ 
as well as the BPS conditions of a (worldline) locally supersymmetric extension of a relativistic, massless particle with moving in the moduli space~${\cal M}$. Indeed, imposing $r'=N$ on the constraint $N^2=r'^2 - r^2 \phi'^\mu g_{\mu\nu} \phi'^\nu$ implied by the $N$-variation of the above action yields
$$
r^2 \phi'^\mu g_{\mu\nu} \phi'^\nu = 0.
$$
Therefore we can reinterpret $r^2=1/e$ as the inverse  einbein of a massless relativistic particle moving in~$\cal M$. The coupling of this particle
to worldline fermions $\theta^i_A=(\theta^*_A,\theta_A^{\phantom{*}})$ is determined by requiring that their supersymmetry variations coincide with those obtained by dimensional reduction
of the four dimensional SUGRA variations. This leads to a one dimensional SUGRA with action principle
$$
S=\int \!dt\ \Big\{
\frac1{2e}\xo^\mu g_{\mu\nu}\xo^\nu
+\frac12\theta^i_A\frac{\nabla \theta_i^A}{dt}
+\frac{\Lambda }{4}\ e\,  \theta^i_{A}  \theta^{\phantom{i}}_{iB} \ \theta^{jA} \theta_j^B
\Big\}\, .
$$
In this formula 
$$
\xo^\mu \equiv \dot x^\mu - V^\mu{}_\alpha^A \theta_A^i \psi_i^\alpha\, ,
$$
is the supercovariantized tangent vector and $\psi_i^\alpha$ are worldline gravitini; the gauge fields for the four
local worldline supersymmetries. The BRST quantization of this supersymmetric spinning particle model is a central focus of this paper.

\section{HyperK\"ahler Sigma Model}

\label{sigma}


We now construct a supersymmetric, non-linear sigma model in a 
  $4n$-dimens\-ional, hyperK\"ahler target space~$(M,g_{\mu\nu})$. 
The field content of the model consists of bosonic worldline embedding coordinates~$
x^\mu(t)
$,
and fermionic spinning degrees of freedom
$
\theta^i_A(t)
$.
Their dynamics are governed by the simple action
\be
S=
\frac12 \int \!dt\ \Big\{  \dot x^\mu g_{\mu\nu}\dot x^\nu
+\theta_A^i \ \frac{\nabla\theta^A_i}{dt}
\Big\}
\, .\label{SHK}
\ee
The (rigid)  symmetries of the model are
\begin{enumerate}
\item {\it Worldline translations:}
\be
\delta x^\mu =\xi \dot x^\mu\, , \qquad \delta \theta^i_A=\xi \dot \theta^i_A\, .
\label{worldline}
\ee
\item {\it $Sp(2)$ $R$-symmetry:} \be
\delta \theta_A^i=\lambda^{ij}\theta_{Aj}\, ,\qquad \lambda^{ij}=\lambda^{ji}\, .\label{R'}
\ee
\item {\it ${\cal N}=4$ supersymmetry:}
\be
\delta x^\mu = V^\mu{}^A_\alpha \theta^i_A \varepsilon^\alpha_i\, ,\qquad
{\cal D} \theta^i_A=-\dot x^\mu V_\mu{}_A^\alpha \varepsilon^i_\alpha\, .
\label{SUSY}
\ee
\end{enumerate}
Here $V^\mu_{~m}=V^\mu_{A\alpha }$ are the inverse vielbeine\footnote{
The vielbeine/orthonormal frames, denoted $V_\mu{}^m$ obey
\be\nn
V^{\phantom{A}}_{\mu}\!{}^A_\alpha\, V^{\phantom{A}}_{\nu}{\!}^\alpha_A=-g_{\mu\nu}\, ,\qquad
V^{\phantom{A}}_{\mu}\!{}^A_\alpha\,  V^\mu{}_B^\beta=-\delta^A_B\delta_\alpha^\beta\, .
\ee
Special holonomy dictates that in addition to these identities for $V^{\phantom{A}}_\mu\!{}^A_\alpha$
(jocularly, the ``zweimalhalbsovielbein'') it is also true that:
$$
V^{\phantom{A}}_{(\mu}{}^A_\alpha V_{\nu)}{}^\beta_A = -\frac12\ g_{\mu\nu} \delta_\alpha^\beta\, ,
\qquad
V^{\phantom{A}}_{(\mu}{}^A_\alpha V_{\nu)}{}_B^\alpha=- \frac 1{2n} g_{\mu\nu} \delta_B^A\, .
$$
}
 written with split flat indices and ${\cal D}$ is the covariant variation: ${\cal D} \theta^i_A\equiv \delta \theta^i_A - \delta x^\mu \Omega_\mu{}_A^B\theta^i_B$.
On functions of~$x^\mu$ it equals $\delta x^\rho \nabla_\rho$;
it obviates the requirement to vary covariantly constant quantities.
In this regard it helps to observe that $\delta={\cal D}$ when varying scalars
(such as the action).

To see explicitly that the action~\eqref{SHK} is supersymmetric, we note the identities
\bea
{\cal D} \dot x^\mu \ \ \  &=& \ \ \frac{\nabla \delta x^\mu}{dt}\, ,\nn\\[2mm]
\Big[{\cal D},\frac{\nabla}{dt}\Big] \theta^A_i  \!\!& = &\delta x^\mu \dot x^\nu R_{\mu\nu}{}^A_B\theta_i^B
= \delta x^{C\alpha}\  \dot x^{D}_{\alpha}\  \Omega^{A}_{BCD}\theta_i^B\, .
\label{varyids}
\eea
Variations linear in fermions cancel by virtue of the first identity, but there are potentially 
cubic fermion terms proportional to $\frac12 \theta^i_A[{\cal D},\frac{\nabla}{dt}]\theta^A_i$.
Using the second identity we see that these vanish since $\Omega_{ABCD}\theta^A_i\theta^B_j\theta_k^C \equiv 0$.

\subsection{Quantization}

\label{hyperquant}

To quantize the model we write it in first order form
\be\nn
S^{(1)}=\int \!dt\ \Big\{
p_\mu\dot x^\mu + \frac12 \theta_A^i \ \dot\theta^A_i
-\frac12 \pi_\mu g^{\mu\nu} \pi_\nu
\Big\}\, ,
\ee 
where $\pi_\mu =p_\mu+ \theta_A^i \Omega_\mu{}_B^A \theta^B_i$, 
and directly impose the canonical commutation relations
dictated by the Darboux form of the first order kinetic terms:
\be
[p_\mu,x^\nu]=-i\delta_\mu^\nu\, ,\qquad
\{\theta^i_A,\theta^j_B\}=-i\epsilon^{ij}J_{AB}\, .\label{bracket}
\ee
We introduce a  Fock representation on a vacuum state $|0\rangle$ as\footnote{The positive definite quantum mechanical inner product for the spinning
degrees of freedom is defined by taking ${\eta^A}^\dagger=\frac{\pa}{\pa \eta^A}$.} 
\bea\nn
\theta_A^i \mapsto  \bv \, \eta_A  \\[2mm]  \,  -i \frac{\pa}{\pa \eta A}  \ev ~,~~
\vspace{1cm} p_\mu |0\rangle = 0 = \frac{\pa}{\pa \eta^A} |0\rangle.
\eea
The fermionic anticommutator~\eqref{bracket} implies
\be\nn
\{\frac{\pa}{\pa \eta^A},\eta^{B}\}=\delta^B_A\, ,
\ee
so the creation operators $\eta^{A}$ produce Fock states which may be identified with sections of the 
 bundle $\wedge E$:
\be
\Gamma(\wedge E)\ni \Phi \equiv \phi_{A_1\ldots A_k}(x)\eta^{A_1}\cdots\eta^{A_k}|0\rangle
\equiv |\phi_{A_1\ldots A_k}\rangle\, .
\label{states}
\ee

The form of  $\pi_\mu$ in the action above may be understood in terms of this representation;
in general 
the covariant momentum is 
\be\nn
\pi_\mu = p_\mu -\frac i2 P_{\mu \,mn}M^{mn}\, ,
\ee
where $M^{mn}$  generate the local Lorentz algebra
\be\nn
[M^{mn},M^{rs}]=M^{ms}\eta^{nr}-M^{ns}\eta^{mr}+M^{nr}\eta^{ms}-M^{mr}\eta^{ns}\, .
\ee
For hyperK\"ahler manifolds the spin connection acts as $P_{mn}M^{mn}= \Omega_{AB} T^{AB} $
where $T^{AB}$, defined in~\eqref{Tt}, generate  $\mathfrak{sp}(2n)$. 
On $\wedge E$ one may alternatively represent $\mathfrak{sp}(2n)$ by bilinears in the spinning degrees of freedom;
\bea \label{TAB}
{T}_{AB} \equiv 
-2\eta_{(A}\frac{\pa}{\pa \eta^{B)} } \, 
\eea
acts identically on $\Phi$ to the operator introduced in~\ref{Tt}. 
This explains the form of $\pi_\mu$; 
acting on $\wedge E$-valued states it produces the covariant derivative\footnote{As usual for first quantized models, $\pi_\mu\pi_\nu\neq \nabla_\mu\nabla_\nu$ because 
$\pi_\mu$ does not see the open index of $\pi_\nu$.} 
\be\nn
\pi_\mu
=p_\mu  -\frac i2\Omega_{\mu AB} {T}^{AB}=-i\nabla_\mu \, .
\ee

\subsection{Charges}

\label{Charges}

Our next task is to write down charges generating the symmetries~\eqref{worldline}-\eqref{SUSY}. 
At the quantum level these are subject to ordering ambiguities which we resolve by relating
symmetry charges and geometric operations. Firstly, we expect the Hamiltonian -- the generator of worldline translations -- to correspond 
to the Laplacian ${\bm \Delta}\equiv \nabla_\mu\nabla^\mu\ $:
\be\nn
-2H \Phi = {\bm \Delta} \Phi\, .
\ee
This is true so long as we adopt the quantum ordering
\be\nn
H=\frac12 \pi_{A\alpha} \pi^{A\alpha} - \frac i2 \Omega_{A\alpha}{}^A_B\pi^{B\alpha}\, ,\qquad
\pi^A_\alpha\equiv V^\mu{}^A_\alpha\pi_\mu\, .
\ee
The four supercharges transform as a doublets under the $\frak{sp}(2)$ holonomy subalgebras
as well as under a Lefschetz-Verbitsky  $\frak{sp}(2)$ algebra which we introduce below.
They are built from the $\frak{sp}(2n)$ contraction of the spinning degrees  of freedom $\theta_A^i$ with the covariant momenta.
On states they act as
\bea\nn
Q^i_\alpha
\equiv \bv  {\bf d}_\a  \\[2mm] {\bm \delta}_\a\ev 
\equiv
\bv   \eta^A \nabla_{\a A}   \\[2mm] -\nabla_{\a}^{ A} \frac{\pa}{\pa \eta^A}  \ev 
\, ,
\eea
where, again, the operator ordering is chosen based on the natural geometric action:
\bea
Q^i_\alpha \Phi 
=  \bv  {\bf d}_\a \\[2mm]   {\bm \delta}_\a  \ev \Phi
 = \bv 
| \nabla_{\alpha[A_1}\phi_{A_2\ldots A_{k+1}]}  \rangle  
\, \nn \\[3mm] 
| k \nabla_{\a A} \phi^A{}_{A_2\ldots A_k} \rangle 
 \ev \, . \nn
\eea
The operator
\be
{\bf d}^\alpha:\Lambda^kE\rightarrow H\otimes \Lambda^{k+1}E\, ,\nn
\ee
belongs to a sequence of Dirac operators introduced by Baston in a study of quaternionic
complexes~\cite{MR1165872}. 
Indeed the operators ${\bf d}_{\alpha}$ and ${\bm \delta}_\alpha$ are analogous to the 
Dolbeault operators on forms, but they act on $\Gamma (\wedge E)$ instead of 
$\Gamma (\wedge TM)$.


%

Next, we present the $R$-symmetry charges generating~\eqref{R'}. 
They can be derived from geometric grounds alone as follows: 
Firstly observe that since we deal with wavefunctions~\eqref{states},
there is no prohibition on adding anti-symmetric $E$-tensors with differing number of indices.
The state $\Phi$ in~\eqref{states} is in fact an eigenstate of the number or ``index'' operator
\be \label{N}
{\bf N}=\eta^{A} \frac{\pa}{\pa \eta^A} \, .
\ee
The invariant tensor $J_{AB}$ allows us to construct two further bilinears,
\be \label{gtr}
{\bf tr}= \frac{\pa}{\pa \eta_A}\frac{\pa}{\pa \eta^A} \, ,\qquad {\bf g}=\eta^{A}\eta_A\, .
\ee 
These act on states as suggested by their names; the operator ${\bf tr}$ removes a pair of indices
by tracing with the invariant tensor $J_{AB}$\ :
\be\nn
{\bf tr}\ |\phi_{A_1\ldots A_k}\rangle = k(k-1)|\phi^{~ ~A}_{A~\,A_3\ldots A_k}\rangle\, .
\ee
Conversely, its adjoint, ${\bf g}$ adds a pair of indices by multiplying by $J_{AB}$ and
antisymmetrizing:
\be\nn
{\bf g}\ |\phi_{A_1\ldots A_k}\rangle = |J_{[A_1A_2}\phi_{A_3\ldots A_{k+2}]}\rangle\, . 
\ee
We arrange these generators in a symmetric $2\times 2$ matrix 
\be \label{fij}
{f}^{ij}=
\left(
\begin{array}{cc}
 {\bf g} & {\bf N}-n \\[2mm]
{\bf N}-n &  {\bf tr}
\end{array}
\right)\ 
\, .
\ee
These are precisely the 
charges corresponding to the $R$-symmetries~\eqref{R'}
and obey the $\mathfrak{sp}(2)$ algebra
\be\nn
[{f}^{ij},{f}^{kl}]=\epsilon^{ki}{f}^{jl}+\epsilon^{kj}{f}^{il}+\epsilon^{li}{f}^{jk}+\epsilon^{lj}{f}^{ik}\, .
\label{ff}
\ee
We note that one may view this representation of $\mathfrak{sp}(2)$ as the Howe dual of the  representation of $\mathfrak{sp}(2n)$ generated by ${T}_{AB}$ 
({\it i.e.}, $\frak{sp}(2)$ and $\frak{sp}(2n)$ are the commutants of one another in $\frak{so}(2n,2n)$). 
In an equation
$$
[{f}^{ij},{T}_{AB}]=0 \, .
$$
Moreover, the  quadratic Casimirs of 
these two  algebras are related by
\be\label{c}
{\bf c}
={\bf g}\ {\bf tr}-{\bf N}({\bf N}-2n-2)
=\frac12 {f}^{ij}{f}_{ij}+n(n+2)
=-\frac12{T}^{AB}{T}_{AB}
 \, .
\ee

The above geometric operators are closely related to the $\mathfrak{so}(4,1)$ Verbitsky algebra acting on differential 
forms on hyperK\"ahler manifolds. (An elegant description of this algebra from a supersymmetric quantum mechanical 
viewpoint is give in~\cite{FigueroaO'Farrill:1997ks}.) In fact $\{\bf g, N, tr\}$ generate an $\mathfrak{sp}(2)$ subalgebra of
$\mathfrak{so}(4,1)$ corresponding to writing $dx^\mu$ as $dx^A_\alpha$ and studying Verbitsky
transformations which do not act on the $H$-index $\alpha$. Alternatively, we may view this 
algebra as a generalization of the  Lefschetz subalgebra that acts on forms on a K\"ahler manifold. Henceforth we
adopt the hybrid designation ``Lefschetz--Verbitsky algebra''.

After some calculation we 
find\footnote{It is interesting to note that this algebra is an In\"on\"u--Wigner contraction of the $\mathfrak{osp}(2|2)$ superalgebra
where the bosonic $\mathfrak{sp}(2)$ and $\mathfrak{so}(1,1)$ blocks are generated by ${f}_{ij}$ and $H$ respectively while $Q_\alpha^i$ belong to off diagonal fermionic blocks. The rescaling of $\mathfrak{osp}(2|2)$ generators $H\rightarrow \lambda^2 H$ and ${Q}_\alpha^i\rightarrow \lambda \, { Q}_\alpha^i$, and the limit $\lambda\rightarrow\infty$ recovers the algebra above.}

\vspace{.4cm}
\begin{tabular}{ccc}
$\{Q_\alpha^i,Q^j_\beta\}=\frac 12\, \epsilon^{ij}\varepsilon_{\alpha\beta} {\bm \Delta}\, ,$
& &
$[{f}^{ij},Q_\alpha^k]=2\epsilon^{k(i} Q^{j)}_\alpha\, ,$
 \\[5mm]
$[{f}^{ij},{f}^{kl}]=\epsilon^{ki}{f}^{jl}+\epsilon^{kj}{f}^{il}+\epsilon^{li}{f}^{jk}+\epsilon^{lj}{f}^{ik}\, ,$
& &
$[{\bm \Delta},{f}^{ij}]=0=[{\bm \Delta},Q_\alpha^i]\, .$
\end{tabular}
\vspace{-.3cm}
\be \label{HKsum}
\ee

\subsection{Summary}

The hyperK\"ahler sigma model presented in this section (and summarized in figure~\ref{picture}) provides a geometric representation of the algebra
\bea\nn
\{ Q_I, Q_J \}=J_{IJ}  \frak{D} \, ,
\eea
with $J$ the invariant rank two tensor of $\frak{so}(2,2)$. This algebra belongs to the family of orthosymplectic algebras for which the BRST detour quantization procedure~\cite{Cherney:2009mf} was developed.

The most general $R$-symmetry of this algebra is $\mathfrak{so}(2,2)$, with generators $R_{IJ}$ acting as 
$$[R_{IJ},Q_K]=2J_{K[I}Q_{J]}\, .$$ 
Upon breaking the index 
$I={}^i_\a$, so that $J_{IJ}=\e_{\a\b}\e^{ij}$, a Howe dual pair of 
$\mathfrak{sp}(2)$ subalgebras generated by
$R^{i}_{(\a\b) i}$ and $R^{\a (i j)}_{\ ~~\a}$ are readily identified. In our hyperK\"ahler sigma model, only the Lefschetz--Verbitsky  $\frak{sp}(2)$  part of the $R$-symmetry algebra acts non-trivially and is identified by $R^{\a (i j)}_{\ ~~\a} \mapsto {f}^{ij}$.

The model we have written down makes sense also on a quaternionic K\"ahler manifold. The geometric interpretations of the charges and wavefunctions
is unaltered. What does change however is the algebra of charges which is no longer a super Lie algebra, but receives deformations from the
non-vanishing $\frak{sp}(2)$ holonomy of a quaternionic K\"ahler manifold. Fortunately however, these deformations produce a first class constraint
algebra. Therefore local, or {\it spinning} particle models can be constructed by gauging supersymmetries. These are the subject of the next section.

\begin{figure}

\vspace{.8cm}
\begin{center}
\!\!\shabox{
\begin{tabular}{ccc}
&{\it Action}&\\[2mm]
&$S=
\frac12 \int\{  \dot x^\mu g_{\mu\nu}\dot x^\nu
+\theta_A^i \ \frac{\nabla\theta^A_i}{dt}
\}$&\\[6mm]
&{\it States}&\\[2mm]
&$\Gamma(\wedge E)\ni \phi_{[A_1\ldots A_k]}$&\\[6mm]
&\!\!\!{\it Charges and Geometry}\!\!\!&\\[2mm]
SUSY&Hamiltonian&$R$-symmetry\\[1mm]
$Q_\alpha^i= \bv {\bf d}_\alpha \\ {\bm \delta}_\alpha\ev$&$-2H={\bm \Delta}$&
\!\!\!\!\!\!\!\!\!\!${f}^{ij}=
\left(
\begin{array}{cc}
{\bf g} & {\bf N}-n \\[2mm]
{\bf N}-n & {\bf tr}
\end{array}
\right)$\\[5mm]
Quaternionic Dirac\!\!\!\!\!\!\!\!&Laplacian&Lefschetz--Verbitsky\\[8mm]
&{\it Algebra}&\\[2mm]
$[{\bf tr},{\bf N}]=2{\bf tr}$&$[{\bf tr},{\bf g}]=4{\bf N} -4n$&$[{\bf N},{\bf g}]=2{\bf g}$\\[3mm]
$[{\bm \delta}_\alpha,{\bf N}]={\bm \delta}_\alpha$&$\{{\bm \delta}_\alpha,{\bf d}_\beta\}=\frac12\varepsilon_{\alpha\beta}{\bm \Delta}$&
$[{\bf N},{\bf d}_\alpha]={\bf d}_\alpha$ \\[3mm]
$[  {\bf tr} , {\bf d}_\alpha ]=2{\bm \delta}_\alpha$&${\bm \Delta}$ central&
$[{\bm \delta}_\alpha , {\bf g} ]=2{\bf d}_\alpha$
\end{tabular}
}
\end{center}

\caption{\label{picture} Geometric data for the quantized hyperK\"ahler sigma model.}
\end{figure}

\section{Quaternionic K\"ahler, ${\cal N}=4$, $d=1$  SUGRA}
\label{QKSUGRA}

Upon replacing the hyperK\"ahler target space with a quaternionic K\"ahler one,
it  is no longer possible to maintain the rigid ${\cal N}=4$ supersymmetry algebra~\eqref{HKsum}.
However, by requiring the algebra to hold only weakly 
we may instead  study local symmetries. 
There are various choices for first class algebras built from the generators $H$, $Q_\alpha^i$ and $f_{ij}$.
Gauging the Hamiltonian $H$ yields a model which is worldline reparameterization independent---generally a 
desirable feature. Local, ${\cal N}=4$, worldline supersymmetry is achieved by gauging the supercharges $Q_\alpha^i$.
Thereafter, one can also consider gauging some combination of $R$ symmetry generators. From a spinning particle perspective gauging $\{H,Q_\alpha^i\}$ and $\{H,Q_\alpha^i,f_{ij}\}$ might seem most natural. 
In general the choice depends on the particular physical or geometric application one has in mind. Also, in general, when quantizing
a first class constraint algebra, one needs to keep in mind what quantization procedure will be employed. Possibly the 
simplest choice is a na\"ive Dirac quantization where one attempts to impose the constraints directly as operator relations
on the physical Hilbert space. Often however, this is not the most interesting choice, and far more can be learned from
a BRST approach. 

In this section we construct the classical spinning particle models corresponding to the $\{H,Q_\alpha^i\}$ and $\{H,Q_\alpha^i,f_{ij}\}$ gaugings. 
In the remainder of the paper, we will be primarily concerned with the BRST quantization of the former of these. In particular we show, motivated by ideas
from higher spin theories, that gauging the only a single
$R$ symmetry generator ${\bf tr}$ within a BRST detour setting produces a
gauge invariant quantum field theoretical model on quaternionic K\"ahler spaces. 

The first step is to introduce Lagrange multipliers (gauge fields) for each constraint
\begin{center}
\begin{tabular}{|ccc|}\hline
&&\\
{\ Constraints\phantom{g}}&&{\ Gauge Fields\ }\\&&\\\hline &&\\$H\approx0$ &&  Lapse $N$\\[1mm]
   $Q^i_\alpha\approx0$ && Gravitini $\psi^\alpha_i$\\&&\\\hline&&\\
 $H\approx0$ &&  Lapse $N$\\[1mm]
     $Q^i_\alpha\approx0$ && Gravitini $\psi^\alpha_i$\\[1mm] 
     ${f}^{ij}\approx0$ && Yang--Mills $A_{ij}$\\[4mm]\hline
\end{tabular}
\end{center}
In this one-dimensional setting, these gauge fields have no dynamics. The charges 
$Q^i_\alpha$ and ${f}^{ij}$
are the same as those of the hyperK\"ahler sigma model in section~\ref{sigma}, while we add curvature corrections to the Hamiltonian~$H$
reflecting that the background is now quaternionic K\"ahler. These are determined by ensuring that the algebra of charges is first class.
Let us give details for each model separately.

\subsection{Rigid Lefschetz--Verbitsky Model}

Gauging only the $Q_\alpha^i$ and $H$ yields a model with rigid Lefschetz--Verbitsky symmetries. Since we work in a quaternionic K\"ahler target space
as described in section~\ref{special} the connection $\nabla$ now
is both $\mathfrak{sp}(2)$ and $\mathfrak{sp}(2n)$-valued. There are two easy methods to compute the 
(second order) action and its symmetries. The first is to start with the sigma model action~\eqref{SHK}
and to proceed using the Noether method, whose first step couples the gravitini to the supersymmetry current/charges $Q^i_\alpha$. This computation is  analogous to the one employed by Bagger 
and Witten~\cite{Bagger:1983tt} to compute matter couplings to ${\cal N}=2$, $d=4$ SUGRA. 
Alternatively, we can begin with a first order action given by the sum of the standard symplectic current
$\int dt \{p_\mu \dot x^\mu +\frac12
\theta^i_A\dot\theta^A_i\}$ and the product of Lagrange multipliers $(N,\psi_i^\alpha)$ with their corresponding constraint.
Thereafter, a Legendre transformation yields the second order action. The results are equivalent and we find
\be
S=\int \!dt\ \Big\{
\frac1{2N}\xo^\mu g_{\mu\nu}\xo^\nu
+\frac12\theta^i_A\frac{\nabla \theta_i^A}{dt}
+\frac{\Lambda N}{4}\ \theta^i_{A}  \theta^{\phantom{i}}_{iB} \ \theta^{jA} \theta_j^B
\Big\}\, ,\label{SUGRAS}
\ee
which enjoys symmetries:
\begin{enumerate}
\item {\it Local worldline reparameterizations:}
\be\nn
\delta x^\mu = \xi \dot x^\mu\, ,\qquad
\delta \theta^i_A=\xi\dot\theta^i_A\, ,\qquad
\delta N=\frac{d(\xi N)}{dt}\, \qquad
\delta \psi^\alpha_i=\frac{d(\xi\psi^\alpha_i)}{dt}\, .
\ee 
\item {\it Rigid $Sp(2)$ $R$-symmetry:} 
\be\nn
\delta \theta_A^i=\lambda^{ij}\theta_{Aj}\, ,\qquad
\delta \psi^i_\alpha=\lambda^{ij}\psi_{\alpha j}\, .
\nn\ee
\item {\it Local ${\cal N}=4$ supersymmetry:}
\bea
\delta x^\mu &=& \ V^\mu{}^A_\alpha \theta^i_A\varepsilon^\alpha_i\, ,\nn\\[4mm]
{\cal D} \theta^i_A &=& -\frac 1N \xo^\mu V_\mu{}_A^\alpha \varepsilon^i_\alpha\, ,\nn\\[4mm]
\delta N&=&\ \psi_\alpha^i\varepsilon_i^\alpha\, ,\nn\\[4mm]
{\cal D}\psi_\alpha^i&=&\frac{\nabla \varepsilon_\alpha^i}{dt}
+\frac{\Lambda N}{2}\ \theta_A^i\theta^A_j\varepsilon^j_\alpha\, .\nn
\eea 
\end{enumerate}
In these formul\ae, ${\cal D}$ is again the covariant variation, but just like the connection $\nabla$,
it too is now $\mathfrak{sp}(2)$ covariant so that, for example, ${\cal D} \psi^i_\alpha = \delta \psi_\alpha^i
-\delta x^\mu \omega_\mu{}_\alpha^\beta\psi^i_\beta$. Also, we have introduced the 
supercovariant tangent vector
\be\nn
\xo^\mu \equiv \dot x^\mu - V^\mu{}_\alpha^A \theta_A^i \psi_i^\alpha\, .
\ee

To verify invariance of this action, notice that the supercovariant tangent vector transforms as
\bea\nn
{\cal D}\!\xo^\mu&=&\frac{\delta N}{2N}\xo^\mu +  V^\mu{}^A_\alpha \Big\{ \frac{\nabla \theta^i_A}{dt}\varepsilon_i^\alpha-\theta^i_A{ \Delta}\psi^\alpha_i\Big\}
+\frac 1N\ {\xo_\nu}\ V^{[\mu}{}^A_\alpha V^{\nu]}{}^\beta_A \varepsilon^i_\beta \psi_i^\alpha\, .
\eea
Here ${ \Delta}\psi^\alpha_i\equiv {\cal D}\psi^i_\alpha - \frac{\nabla \varepsilon_\alpha^i}{dt}$ is shorthand for the
two fermion gravitini variations. The last terms are of the form $\xo_\nu A^{[\mu\nu]}$ so do not contribute to
the variation of the bosonic matter kinetic term $\frac1{2N}\xo^2$ while the leading term 
perfectly ensures the kinetic terms vary into
\be
\delta\int\Big\{\frac1{2N}\xo_\mu \!\xo^\mu+\frac12\theta^i_A\frac{\nabla\theta^A_i}{dt}\Big\}=
\int\Big[
-\frac1N\xo_A^\alpha{ \Delta}\psi_\alpha^i+\frac12\theta^i_A\Big[{\cal D},\frac{\nabla}{dt}\Big]\Big]\theta^A_i
\, .
\ee
These cancel the variation of the four point fermi coupling to the Riemann tensor. This
relies on the quaternionic K\"ahler analog of the identity~\eqref{varyids} which yields $\delta x\, \dot x$
times the Riemann tensor for the commutator of covariant worldline derivatives and variations.
Trading $\dot x$ for $\xo$ yields exactly the terms required to cancel the variation of the lapse $N$ 
multiplying the four point coupling.

A final point worth stressing is that the parameter $\Lambda$ is not fixed by the requirement
of local supersymmetry in one dimension. In dimension four, coupling ${\cal N}=2$ SUGRA to 
matter fixes the scalar curvature in terms of Newton's constant~${\kappa}$~\cite{Bagger:1983tt}
(This follows
by requiring variations of the Einstein--Hilbert and Rarita--Schwinger terms to cancel at
order $\kappa^0$ in the Noether procedure.) Both these terms are absent in our one dimensional
model.  

\subsection{Gauged Lefschetz--Verbitsky Model}
To gauge the Lefschetz--Verbitsky $\mathfrak{sp}(2)$ symmetry 
we need only replace the covariant derivative $\nabla$ in~\eqref{SUGRAS}
by its $\mathfrak{sp}(2)$ covariantization $\stackrel{\rm A}{\nabla}$ defined by
\be\nn
\frac{\stackrel{\rm A}\nabla\! v^i}{dt} \equiv \frac{\nabla v^i}{dt}+ A^{ij} v_j\, .
\ee
Therefore the gauged action reads
\be
S=\int \!dt\ \Big\{
\frac1{2N}\xo^\mu g_{\mu\nu}\xo^\nu
+\frac12\theta^i_A\frac{\stackrel{\rm A}{\nabla} \!\theta_i^A}{dt}
+\frac{\Lambda N}{4}\ \theta^i_{A}  \theta^{\phantom{i}}_{iB} \ \theta^{jA} \theta_j^B
\Big\}\, ,\label{SUGRASG}
\ee
which differs from~\eqref{SUGRAS} by a Lagrange multiplier term
$\int \frac12 \theta^i_A A_{ij} \theta^{jA}$ (so the gauge field $A_{ij}$ is
a unit weight, worldline tensor density or volume form). In addition to the new local
Lefschetz--Verbitsky symmetry
\be\nn
\delta \theta^i_A=\lambda^i_j \theta^j_A\, ,\qquad
\delta \psi^i_\alpha= \lambda^i_j \psi^j_\alpha\, ,\qquad
\delta A^{ij}=\dot \lambda^{ij}+2A^{k(i}\lambda_k^{j)}\, ,
\ee
the supersymmetry transformations are modified to read
\bea
\delta x^\mu &=& \ V^\mu{}^A_\alpha \theta^i_A\varepsilon^\alpha_i\, ,\nn\\[4mm]
{\cal D} \theta^i_A &=& -\frac 1N \xo^\mu V_\mu{}_A^\alpha \varepsilon^i_\alpha\, ,\nn\\[4mm]
\delta N\ &=&\ \ \psi_\alpha^i\varepsilon_i^\alpha\, ,\nn\\[4mm]
{\cal D}\psi_\alpha^i&=&\frac{\stackrel{\rm A}{\nabla}\! \varepsilon_\alpha^i}{dt}
+\frac{\Lambda N}{2}\ \theta_A^i\theta^A_j\varepsilon^j_\alpha\, ,\nn\\[4mm]
\delta A^{ij}&=&\ \ 0\, .\nn
\eea 
These results and other gaugings follow easily from the canonical analysis of
the next section.

\subsection{Dirac Quantization}

To perform a canonical analysis and Dirac quantization of the 
rigid Lefschetz--Verbitsky model we first note that the symplectic structure $\int \!dt\ \Big\{
p_\mu\dot x^\mu + \frac12 \theta_A^i \dot\theta^A_i\}$ implies the same Fock space structure as in the hyperK\"ahler
case (see in particular formul\ae~(\ref{bracket}-\ref{states})). The Dirac Hilbert space is therefore again
sections of the antisymmetric $\mathfrak{sp}(2n)$ 
tensor bundle $\wedge E$.

The (quantized) supercharges $Q_\alpha^i$ and Lefschetz--Verbitsky generators take the same form 
as in the analysis of the hyperK\"ahler sigma model in section~\ref{Charges}. The Hamiltonian~$H$ receives a
curvature correction term (implied by the four-fermi term in the action~\eqref{SUGRAS}  proportional to the lapse~$N$)
Again these charges may all be quantized with orderings obtained by ensuring that the quantum algebra of constraints is first class.
The Dirac quantization of the model then amounts simply to imposing the conditions $H\Psi=Q_\alpha^i\Psi=0$ 
on wavefunctions $\Psi$ valued in $\Gamma(\wedge E)$. (The gauged Lefschetz--Verbitsky model incurs the additional
constraint $f_{ij}\Psi=0$.) We pay little attention to an analysis of this quantum system because it suffers a certain deficiency
which we now explain, and will remedy in the next section by means of a BRST analysis:

On a quaternionic K\"ahler manifold we must remember that the spin connection has both  $\frak{sp}(2n)$ and $\frak{sp}(2)$
valued parts which couple naturally to the respective generators $T_{AB}$ and $t_{\alpha\beta}$. However, from the spinning degrees of freedom $\theta_A^i$  of this model, we can {\it only} build a representation of the $\frak{sp}(2n)$ generators $T_{AB}$. On the one hand, this seems sufficient
because acting on $\wedge E$-sections, we still have $i\pi_\mu=\nabla_\mu$. 
But, acting with a supersymmetry generator $Q_\alpha^i$ introduces an $\frak{sp}(2)$ index~$\alpha$, and we seem to have no way, in the
spinning particle model context, to obtain further covariant derivatives acting correctly on~$\alpha$. A geometer might consider constructing
supersymmetry-like operators built from the covariant derivative by {\it fiat} (and in fact, the geometric calculus section~\ref{calculus} of this paper
can be taken on its own and read this way). However, there is a very natural physical mechanism to introduce additional spinning degrees of freedom 
that can represent the $\frak{sp}(2)$ generators $t_{\alpha\beta}$. In fact, this is precisely what BRST quantization of the model does.

\section{BRST and the Geometry of Ghosts}

\label{BRST}

The one dimensional quaternionic K\"ahler spinning particle model enjoys local worldline 
supersymmetry and reparameterization invariances. This implies that they form a first class algebra (even though the supercharges 
do not commute with the Hamiltonian unlike those in the hyperK\"ahler sigma model where they generate genuine symmetries). 
In this section we present the nilpotent, quantum,  BRST charge for this algebra. 
Again, unlike the hyperK\"ahler model, this constraint algebra is higher rank; it does not form a Lie algebra. 
This means that, in principle,
we need to resort to homological perturbation methods to construct the BRST charge. (The reader may consult~\cite{Teitelboim} for 
a detailed account of the analysis of gauge theories using BRST techniques and in particular the construction of a nilpotent BRST charge for higher rank algebras.) 
Although standard, such a computation is 
rather involved, so instead we present a solution relying on the underlying quaternionic geometry.

The general structure of the BRST charge we search for is given by expanding it in powers of the worldline reparameterization ghost $c$ and its antighost
$b$ represented as $\frac{\partial}{\partial c}$
\be
Q_{\rm BRST}=c\, \frak{D} + {\cal Q} - M \frac{\partial}{\partial c}\, \, .\label{QBRST}
\ee
If our constraint algebra were a Lie algebra (as it is in the hyperK\"ahler case), the operator ${\cal D}$ would be the worldline Hamiltonian and ${\cal Q}$ the contraction of the supercharges with commuting supersymmetry ghosts $c^\alpha_i$. 
However, since we have a higher rank constraint algebra,
we must add terms with higher powers of ghosts and antighosts. 
We determine these by making a simple geometric {\it ansatz} for 
${\cal Q}$ and then requiring nilpotency of~$Q_{\rm BRST}$. 

The key geometric idea is that ghosts and antighosts can be used to represent the $\frak{sp}(2)$ special holonomy generators. The quantized commuting superghosts $c_i^\alpha$ and superantighosts~$b^i_\alpha$ with algebra
\be
[b^i_\a, c^\b_j]=\delta^i_j \delta^\b_\a\label{ghalg}
\ee 
allow formation of bilinears
$c^i_\alpha b^j_\beta- c^j_\beta b^i_\alpha$ 
that generate a faithful representation of
$\frak{so}(2,2)$,
the $R$-symmetry algebra of our first class constraint superalgebra, on the ghosts (and/or antighosts). 
Specializing to the Howe dual subalgebras generated by
\bea\nn
f_{ij}^{\rm gh}&=&\,-2 c_{(i}^\alpha b_{j)\alpha}^{\phantom{\alpha}}\, 
,\nn\\[2mm]
t_{\rm gh}^{\alpha\beta}&=&-2c^{ i (\alpha } b^{\beta)}_{\, i} \, ,
\eea
we obtain representations of the Lefschetz--Verbitsky  and $H$-bundle special holonomy $\frak{sp}(2)$ algebras, respectively. (We will discuss the precise definition of the superghost Hilbert space at the end of this section, but for now concentrate on building a nilpotent BRST charge.)

This means that we can solve the problem of the covariant momentum operator $\pi_\mu$ discussed in the previous section---namely that it was not covariantized with respect to the sp(2) holonomy---by using the above ghost representation for $t_{\alpha\beta}$. So we now construct a covariant momentum operator
\be
\Pi_\mu \equiv p_\mu - \frac i2 \Omega_\mu{}^A_B T^B_A - \frac i2 \omega_\mu{}^\alpha_\beta t^\beta_\alpha\, , \label{covmom}
\ee
which acts on both $E$ and $H$ bundles. (In some sense, the ghosts play the {\it r\^ole} of frames for the bundle~$H$.) In turn we introduce BRST-extended supersymmetry charges
$
\theta^i_A V^\mu{}^A_\alpha \Pi_\mu\
$
and consider the ansatz
\bea \nn
{ \cal Q} \equiv i c^\alpha_i  
\bv \eta^A \\[2mm] \frac{\pa }{ \pa \eta_A} \ev^{\! i}
 V^\mu{}_{\alpha A} \Pi_\mu\, .
\eea
for the form of equation \eqref{QBRST}. 

Before proceeding, it is worth noting that we have actually found a new Dirac operator: 
Reunifying $\mathfrak{sp}(2)$ and $\mathfrak{sp}(2n)$ indices as a single
$\mathfrak{so}(2n,2n)$ index $m=A\alpha$ and forming the combination
\bea\nn
\gamma^m=c^{ \alpha i} \bv \eta^A \\[2mm] \frac{\pa }{ \pa \eta_A} \ev_{\! i} \, ,
\eea
we find a Clifford algebra
$$
\{\gamma^m,\gamma^n\}= M \, \eta^{mn}\, ,
$$
$$
M\equiv \frac12  c^{\alpha i} c_{\alpha i} \, .
$$
Since the covariant momentum~\eqref{covmom} acts as the covariant derivative, a Dirac-type operator follows 
\be\label{DIRAC}
{\cal Q}=\gamma^m \nabla_m \, .
\ee

Returning to our BRST charge computation, a simple Weitzenbock-like calculation\footnote{Note that the computation of the term coupling the 
curvature  to two Dirac matrices relies heavily on $\gamma^m$ being a composite built from ghosts
and spinning degrees of freedom.} shows
\be
{\cal Q}^2 = 
M \frak{D} \, ,\label{QQSUSY}
\ee
where the BRST-extended Hamiltonian is 
\be\nn
2 \, \frak{D}=\square - \frac14 (f^{ij} + f^{ij}_{\rm gh} )(f_{ij} + f_{ij}^{\rm gh}) - \frac n2(n+2) \, .
\ee
In this expression,  
$\square= \Delta +\frac14(T^2 + t^2)$
is a 
 quaternionic K\"ahler Lichnerowicz wave operator,
which will be introduced in Section~\ref{calculus}. It  satisfies $[\square , {\cal Q}]=0$.
Further,  since $f^{ij}$ and $f_{\rm gh}^{ij}$ obey
$[f_{ij}+f^{\rm gh}_{ij}, c^\a_k \,Q^k_\a]=0$ 
and the latter 
commutes with\footnote{In fact, 
linear combinations of the ghost bilinears mentioned below equation~\eqref{ghalg} 
are precisely those which commute with $M$. }  $M$ , we have the following identities
\be
[\frak{D},M]=[{\cal Q} \,,\frak{D}]=[{\cal Q} \, ,M]={\cal Q}^2-M\frak{D}=0 \, .
\ee
These immediately imply that the BRST charge~\eqref{QBRST} is nilpotent.
The form of this BRST charge is exactly suited to the detour quantization methods
of~\cite{Cherney:2009mf}. 
To that end we next specify our choice of ghost vacuum. 

We represent the ghost algebra~\eqref{ghalg}  in a Fock representation 
by splitting the ghosts and antighosts 
into derivatives and power series coordinate coefficients.
The choice of vacuum is determined by
splitting the Verbitsky--Lefschetz doublets as
\bea \label{detour polarization}
c_i^\alpha=\left(\begin{array}{cc } z^\alpha & \frac{\pa}{\pa p_\a}\end{array} \right)\, ,\qquad
b_i^\alpha=\left(\begin{array}{cc} -p^\a & \frac{\pa}{\pa z_\alpha} \end{array}\right)\, .
\eea
Therefore we may view $(z^{\alpha},p^{\alpha})$ as creation operators for symmetric $H$-bundle indices. So states $\Phi$ in the superghost extended Hilbert space are sections of
$$
\Gamma(\wedge E\otimes (\odot H)^{\otimes 2})\ni \Phi
\equiv \phi_{A_1\ldots A_k}{}_{\alpha_1\ldots \alpha_s}^{\beta_1\ldots \beta_t}{\scriptstyle (x)}\,
\eta^{A_1}\cdots\eta^{A_k}z^{\alpha_1}\cdots z^{\alpha_s}
p_{\beta_1}\cdots p_{\beta_t}|0\rangle
$$
\be
\hspace{1.9cm}
=\ |\phi_{[A_1\ldots A_k]}{}_{(\alpha_1\ldots \alpha_s)}^{(\beta_1\ldots \beta_t)}\rangle
\ =\ 
\Phi_{\tiny\Yvcentermath1 
\!\!\!\mbox{$k$}
\left\{
\yng(1,1,1,1,1)
\right.
\otimes 
\begin{array}{c}
\overbrace{\yng(6)}^{t}\\
\otimes\\
\underbrace{\yng(4)}_{s}
\end{array}
}\, .
\nn
\ee
In the Young diagram notation the column denotes antisymmetrized $E$-indices while 
the rows are symmetrized $H$-indices.

We now have a well-defined BRST cohomology. Before analyzing it via BRST detour methods, we take a short geometric excursion to develop a quaternionic calculus of the various operators that will appear in those results.

\section{A Quaternionic Geometric Calculus}

\label{calculus}
On a $d$-dimensional Einstein manifold the Riemann tensor decomposes as
\be\nn
R_{\mu\nu\rho\sigma}=\underbrace{\frac{2\Lambda}{(d-1)(d-2)} (g_{\mu\rho}g_{\nu\sigma}-g_{\nu\rho}g_{\mu\sigma})}_{\rm Constant \  Curvature} \ + \ 
\underbrace{W_{\mu\nu\rho\sigma}\phantom{\!\!\!\!\!\!\!\frac1{(2)}}}_{\rm Weyl} .
\ee
The special constant curvature case---when the Weyl tensor vanishes---enjoys many distinguishing properties, including a Lichnerowicz wave operator
which commutes with generalized gradient and divergence operators acting on tensors of very general types. Comparing this formula with the one
for the quaternionic K\"ahler Riemann tensor in~\eqref{Riemann} we see that the totally symmetric tensor $\Omega_{ABCD}$ plays a {\it r\^ole}
similar to the Weyl tensor\footnote{In fact, in four dimensions it plays the {\it r\^ole} of the anti-self dual Weyl tensor~\cite{Salamon,MR1165872}.}; 
if we could somehow find a ``regime'' in which it did not contribute we might be able to analyze quaternionic K\"ahler
geometry along lines similar to the constant curvature case.

In fact, exactly such a regime does exist, namely sections of the product of $\wedge E$ with the tensor bundle ${\cal T} H$
(with sections being arbitrary $H$-tensors)
\be\nn
\Gamma(\Lambda E\otimes {\cal T} H)\ni \phi_{[A_1\ldots A_k]}{}^{\alpha_1\ldots \alpha_s}\, ,
\ee
the idea being that antisymmetry in $\frak{sp}(2n)$ indices prevents the totally symmetric tensor $\Omega_{ABCD}$ from contributing.

In particular, the central operations will be the quaternionic generalizations of the Dolbeault operators
\be\nn
\begin{array}{rccc} 
\left(\begin{array}{c} \DI^\alpha \\ \DE^{\alpha}\end{array}\right):&
 \Gamma(\Lambda E\otimes {\cal T} H)&\longrightarrow& \Gamma(\Lambda E\otimes {\cal T} H)^{\otimes2}\\
&\rotatebox{90}{$\in$}&&\rotatebox{90}{$\in$}\\
&\phi_{[A_1\ldots A_k]}{}^{\alpha_1\ldots \alpha_s}&\mapsto&
\left(\begin{array}{c}\nabla^\alpha_{[A_1} \phi^{\phantom{alpha}}_{A_2\ldots A_{k+1}]}{}^{\alpha_1\ldots \alpha_s} \\[2mm]  
k\nabla^{\alpha }_A\,  \phi^A{}_{[A_2\ldots A_k]}{}^{\alpha_1\ldots \alpha_s}\end{array}\right)
 \end{array}
\ee
These operators are motivated by the quantized supersymmetry charges of the previous sections, but are more general since they can act on
arbitrary $H$-tensors. For computations, it is often useful to adopt a hybrid $E$-index free notation where
\bea\nn
\phi_{[A_1\ldots A_k]}{}^{\alpha_1\ldots \alpha_s} &\to& \Phi^{\alpha_1\ldots\alpha_s} = \phi_{A_1\ldots A_k}{}^{\alpha_1\ldots \alpha_s}\eta^{A_1}\cdots \eta^{A_k}\, ,\\[3mm]
{\bf d}^\alpha &=& \eta^A\nabla^\alpha_A\, ,\nn\\[1mm]
{\bm \delta}^\alpha &=& - \nabla^{\alpha A}\frac{\partial}{\partial \eta^A}\, ,\nn
\eea
and the Grassmann variables $\eta^A$ play the {\it r\^ole} of the anticommuting differentials $dx^\mu$ employed in the theory of differential forms. 

The non-dynamical Lefschetz--Verbitsky charges
\be\nn
{f}^{ij}=\left(\begin{array}{cc} \GI&{\bf N}-n\\[3mm]{\bf N}-n&\TR\end{array}\right)
\ee
act exactly as described in~\ref{Charges} on the antisymmetric $E$-indices (with the same expressions in terms of $\eta$'s), namely adding or removing pairs of antisymmetrized indices
using the invariant tensor $J_{AB}$ or counting indices.
In terms of these
$\DI^\alpha$, $\DE^{\alpha}$ obey a very elegant algebra
\bea\label{qalg}
&\{\DI^\alpha,\DI^{\beta}\}=- \frac12 \GI~ t^{\alpha\beta}\, ,&\nonumber \\[3mm]
&\{\DI^{\alpha} , \DE^\beta\}=  \frac 1 2 \varepsilon^{\alpha \b}({\bm \Delta} - {\bf c}) 
- \frac12 t^{\alpha\beta}({\bf N}-n)\, ,&\nonumber\\[3mm]
&\{\DE^\alpha,\DE^{\beta}\}   =  -  \frac12 \TR~ t^{\alpha\beta}~,&
\eea 
where ${\bf c}$ is again the Lefschetz--Verbitsky $\frak{sp}(2)$ Casimir operator of~\eqref{c}. 

These formul\ae~can be repackaged even more simply by noticing that the operator 
\be\nn
~~~~~~\square={\bf \Delta}+\frac 1 4 T^{2}+\frac 1 4 t^{2}~,~~~~~~~~~~~~~\textrm{with}~~\left\{\begin{array}{rcl} T^2&=&T_{AB}T^{AB}\\[2mm]t^2&=&t_{\alpha\beta}t^{\alpha\beta}\end{array}\right.
\ee
{\it commutes} with ${\bf d}^\alpha$ and ${\bm \delta}^\alpha$. This is an extremely important result, so we shall call $\square$ a 
{\it quaternionic K\"ahler Lichnerowicz wave operator}. Its existence validates our claim that by studying the bundle $\wedge E\otimes {\cal T} H$,
quaternionic K\"ahler geometry could be made to mimic its constant curvature counterpart.

Specialized to totally symmetric $H$-tensors, the operators $({\bf d}^\alpha,{\bm \delta}^\alpha)$ coincide with the action of the BRST-extended supersymmetry
charges in section~\ref{BRST}, therefore we adopt the suggestive notation 
\be\nn
{ \cal Q}^i_\alpha=\left(\begin{array}{c}\DI_{\alpha}\\ \DE_{\alpha}\end{array}\right)\, .
\ee 
and call these operators {\it generalized supercharges}.
We may now unify the algebra~\eqref{qalg} as
\be\nn
\{{\cal Q}^i_\alpha,{\cal Q}^j_\beta\}
=\frac 12\,  \varepsilon_{\alpha \b }\epsilon^{ij}\wt{ \square}  
-  \frac12 f^{ij}\, t_{\alpha\beta}~.
\ee
with
$$
\wt{ \square}\equiv
\square 
- \frac 1 4 f_{ij}f^{ij}-\frac 1 4 t_{\alpha\beta}t^{\alpha\beta} -\frac n2(n+2)\, .
$$
It is interesting to note that these formul\ae~ enjoy a complete symmetry when all $H$-indices $\alpha,\beta,\ldots$ are exchanged with 
their Lefschetz--Verbitsky counterparts $i,j,\ldots$. This symmetry  appears more starkly when we compute the products of generalized supercharges
\be\nn
{\cal Q}^{i}_{\alpha}\, {\cal Q}^j_\beta
=\frac 14 \,  \varepsilon_{\alpha \b}\, \epsilon^{ij}\, \wt{ \square}
- \frac14  f^{ij}\, t_{\alpha\beta}
-\frac 12\,  \varepsilon_{\alpha\b}{\bf b}^{ij}
-\frac 1 2 \epsilon^{ij}{\bf b}_{\alpha\beta}\, ,
\ee
where we have defined the bilinears
\be\nn
{\bf b}^{ij} 
\equiv 
{\cal Q}^{ (i}_\a {\cal Q}^{j) \alpha} \, 
, \qquad {\bf b}_{\alpha\beta}
\equiv 
{\cal Q}_{i (\alpha}{\cal Q}_{\beta)}^{\,i} \, .\ee
Observe that, since the generalized supercharges form $\frak{sp}(2)$ doublets under Lefschetz--Verbitsky and $H$-symmetries
\be\nn
[f^{ij},{\cal Q}^k_\alpha]
=\epsilon^{ki}{\cal Q}^j_\alpha 
+ \epsilon^{kj}{\cal Q}^i_\alpha\, ,\qquad
[t_{\alpha\beta},{\cal Q}_\gamma^i]
=\varepsilon_{\gamma\alpha}{\cal Q}_\beta^i + \varepsilon_{\gamma\beta}{\cal Q}_\alpha^i\, ,
\ee
the six charge bilinears ${\bf b}_{\alpha\beta}$ and ${\bf b}_{ij}$ form two adjoint 
$\frak{sp}(2)$ triplets.
This leads one to wonder whether these operators form a pair of $\frak{sp}(2)$ algebras when commuted among themselves.
This question is particularly pressing when we observe that the operator
\be\nn
{\bf d}_\alpha {\bf d}^\alpha + {\bf g}\, ,
\ee
coincides with that introduced by Baston in his construction of quaternionic analogues of Dolbeault cohomology on quaternionic 
K\"ahler manifolds. In fact, this operator is one of a triplet of operators
\be\nn
{\bf B}_{ij}={\bf b}_{ij}+f_{ij}
\ee
which we shall call {\it Baston operators}.  In fact, this structure of $R$-symmetry groups represented in terms of bilinears in supercharges has appeared before~\cite{MR1165872}. For example, for differential forms on a K\"ahler manifold, bilinears in the Dolbeault operators $\{{\bm \delta} \bar{\bm \delta}\, , \bm\Delta -2\bm \partial \bm \delta -2\bar{\bm \partial} \bar{\bm \delta}\, ,\bm \partial \bar{\bm \partial}\}$ obey an $\frak{sp}(2)$ Lie algebra (up to an overall factor of the central 
form Laplacian on the right hand side of commutators). In fact a similar phenomenon holds for more general orthosymplectic algebras~\cite{Cherney:2009md}.
Moreover, the K\"ahler result immediately implies the same algebra for  the ${\bf b}_{ij}$ on hyperK\"ahler manifolds.
In the more general quaternionic K\"ahler case one no longer finds a Lie algebra built from ${\bf b}_{ij}$ but instead the following
rather interesting deformation thereof\footnote{It would be interesting
to investigate whether the last terms in this formula can be absorbed by replacing the
operator $\wt \square$ with the BRST Hamiltonian. Of course, this could only be
the case specializing to the BRST superghost Hilbert space of the previous section.}
\be\nn
[{\bf B}^{ij},{\bf B}^{kl}] =  \epsilon^{(i(k} \Big[
(\wt \square-2)\,  {\bf B}^{j)l)}
+{\bf B}^{j)l)} (\wt \square-2)  
-
f^{j)l)}({\bf b}_{\a\b}t^{\a\b} + \frac12 \, t^2)\Big]
\, .
\ee
The Weyl ordering on the right hand side is necessary because  (as opposed to the quaternionic K\"ahler Lichnerowicz wave operator $\square$) 
the operator $\wt \square$ is not central. Note that the operators ${\bf b}_{\alpha\beta}+t_{\alpha\beta}$ obey an analogous algebra, thanks to the 
aforementioned symmetry between $H$-indices and Lefschetz--Verbitsky ones. The main formul\ae~of this section are summarized in figure~\ref{qcalculus}.
We now orchestrate these geometric results with our BRST detour techniques to construct our main result, a gauge invariant quaternionic K\"ahler
quantum field theory.

\begin{figure}
\begin{center}
\shabox{
\begin{tabular}{c}
{\it Quaternionic Dolbeault Operators}\\[2mm]
$ { \cal Q}^i_\alpha=\left(\begin{array}{c}\DI_{\alpha}\\ \DE_{\alpha}\end{array}\right) $\\[8mm]
{\it Quaternionic Dolbeault Algebra}\\[2mm]
$\{{\cal Q}_i^\alpha,{\cal Q}_j^\beta\}
=\frac 12\,  \varepsilon^{\alpha \b }\epsilon_{ij}\wt{ \square}
-  \frac12f_{ij}\, t^{\alpha\beta}$\\[5mm]
{\it Quaternionic K\"ahler Lichnerowicz wave operator}\\[2mm] 
$\square={\bf \Delta}+ \frac14 (T^2 + t^{2}) 
=\wt\square + \frac 1 4 f^2 + \frac 1 4 t^2 +\frac n2(n+2)$
\\[5mm]
{\it Baston operators}\\[2mm]
${\bf B}^{ij}=Q^{(i}_\a Q^{j) \a} +  f^{ij}$ \\[4mm] 
$
=\left(\begin{array}{cc}
{\bf d}_\alpha {\bf d}^\alpha
+\GI&
\DI_\alpha\DE^\alpha+\DE_\alpha\DI^\alpha+2({\bf N}-n)
\\[3mm]
\DI_\alpha\DE^\alpha+\DE_\alpha\DI^\alpha+2({\bf N}-n)&
\DE_\alpha\DE^\alpha+\TR
\end{array}\right)
$
\\[8mm]
{\it Baston Algebra}\\
$[{\bf B}_{ij},{\bf B}_{kl}] =  \epsilon_{(i(k} \Big[{\bf B}_{j)l)} \wt \square + \wt \square\,  {\bf B}_{j)l)}\Big]$
\\[5mm]
$[{f}^{ij},{\bf B}^{km}]=2 \, \epsilon^{k(j}{\bf B}^{i)m}+2 \, \epsilon^{m(j}{\bf B}^{i)k}$
\end{tabular}
}
\end{center}
\caption{The quaternionic K\"ahler  calculus\label{qcalculus}}
\end{figure}

\newpage

\section{The Quaternionic K\"ahler Detour Complex}

\label{QKdetour}

The BRST detour quantization formalism presented in~\cite{Cherney:2009mf}, takes as its input a BRST charge of the form \eqref{QBRST}, 
together with a representation of the  underlying constraint algebra acting on sections of a bundle over some manifold $M$, and outputs a classical field theory on $M$. The equation of motion, gauge invariances, and Bianchi identities are concisely summarized in a detour complex 
\be \nn
\begin{array}{c}
\cdots \stackrel{\cal{Q}} {\longrightarrow}\left(\!\!{\tiny \begin{array}{c} \mbox{Gauge}\\[1mm]\mbox{parameters}\end{array} }\!\!\right)
\stackrel{ \cal{Q} } {\longrightarrow}
\left(\!\!{\tiny \begin{array}{c} \mbox{Gauge}\\[1mm]\mbox{fields}\end{array}}\!\!\right)
~~~~~~
\left(\!\!{\tiny \begin{array}{c}\mbox{Equations of }\\[1mm]\mbox{motion/currents}\end{array}}\!\!\right)
\stackrel{ \cal{Q} }{\longrightarrow} 
\left(\!\!{\tiny \begin{array}{c}\mbox{Bianchi/Noether}\\[1mm] \mbox{identities}\end{array}}\!\!\right)\stackrel{\cal{Q} }{\longrightarrow} \cdots~~
\\[1mm]
\hspace{-1.3cm} \Big|\hspace{-.7mm}\raisebox{-2.2mm}{\underline{\quad \ \  
\raisebox{-.3mm}{${\scriptstyle {{\cal D}- {\cal Q} M^{-1} {\cal Q} }}$}\quad\;}} 
\hspace{-1.2mm}{\Big\uparrow}
\end{array}
\ee
The $\cdots$ on the ends of the complex describe any gauge for gauge symmetries and their accompanying Bianchi for Bianchi identities.

The  models described by the above complex, depend on towers of gauge fields (possibly infinitely many for the case when the constraint
algebra contains Grassmann odd generators). There are cases when these towers of gauge fields have a simple geometric interpretation
(including the quaternionic K\"ahler models described here--see our conclusions for a discussion of this point). These towers of gauge fields 
arise because the physical cohomology retains a dependence on certain bilinears in ghosts. Generically it is desirable to remove
this ghost dependence; this can be achieved by gauging further combinations of $R$ symmetries (the ``ghostbusting'' procedure of~\cite{Cherney:2009mf}).
This leads to more standard physical models with equations of motion and local invariances of the form
\be\nn
({\bm \Delta} + \cdots) A = 0\, ,\qquad \delta A = D \alpha\, ,
\ee
where ${\bm \Delta}$ is typically the Laplace operator, $A$ denotes  some type of gauge field, and the operator $D$ generates its gauge invariance.
The $\cdots$'s stand for  terms required for the equation of motion to be gauge invariant. The operator ${\bm \Delta} + \cdots$ can be expressed
in a simple ``Labastida'' form (a name which refers to its origin in the theory of higher spin theories) or equivalently as a self-adjoint ``Einstein operator''
(this name was chosen since the linearized Einstein tensor is one of the simplest examples). The latter form immediately implies a gauge invariant  action principle. Let us now apply these results to the model at hand, we focus on the main formul\ae, referring the reader to the articles~\cite{Cherney:2009mf} for detailed derivations of the underlying methodology.

Firstly the ``long operator'' ${\cal D}-{\cal Q} M^{-1} {\cal Q}$ can be defined as acting on wavefunctions 
\be\nn
\Psi(y) \in \wedge E[y]
\ee
built from polynomials in a commuting bilinear in superghosts $y=2z^{\alpha}p_{\alpha}$ with coefficients in $\Gamma(\wedge E)$
(because this space forms the ghost number zero kernel of the operator $M$).
Explicitly it yields a gauge invariant equation of motion
\be\label{LONG}
{\bf B}^{ij}f^{\rm gh}_{ij}\, \Psi = 0\, ,
\ee
where, acting on functions of only $y$, the operators $f_{ij}^{\rm gh}$ have the simple expression
\be\nn
f_{ij}^{\rm gh} = \begin{pmatrix}~y~ &~-2(y\partial_y+1)~\\[4mm]-2( y\partial_y+1)~&~4(y\partial^2_y+2 \partial_y)~\end{pmatrix}\, .
\ee
This model is but a stepping stone to our theory of interest, obtained by also gauging the Lefschetz--Verbitsky generator~${\bf tr}$.
This choice may seem {\it ad hoc}, but is well known  in the higher spin literature (for example, it is necessary to obtain the linearized
Einstein tensor in the case of a spin~2 theory). In particular it removes all dependence of the physical cohomology on the ghost bilinear $y$.
The physical gauge fields now take values in $\wedge E$ only.

In fact, gauging the $R$-symmetry $\TR$ 
amounts to restricting the $y$ dependence of $\Psi(y)$ in the detour complex to 
\be\nn\label{BESSELR}
\Psi=\frac{I_{1}(\sqrt{y \TR})}{\sqrt{y\TR }}\, \varphi \, ,~~ ~~\varphi\in \wedge E  \, ,
\ee
and pushing the long operator in~\eqref{LONG} past the operator-valued Bessel function yields the very simple
``Labastida'' equation of motion
\be\label{Labastida}
\shabox{$\TR\Big(\DI_\alpha\DI^\alpha{}+ 
\GI \Big)\varphi=0$\, .}  \ee
In particular, notice that this equation factorizes as the product of $\TR$ with the operator discovered long ago by Baston~\cite{MR1165872} . 
In fact this gauge theory, on a quaternionic K\"ahler manifold 
mimics the higher form $(p,q)$-form K\"ahler Electromagnetism theory presented in~\cite{Cherney:2009vg} (observe the correspondence between the Dolbeault bilinear ${\bm \pa} \bar{\bm \pa}$ and the Baston operator
${\bf d}_\alpha {\bf d}^\alpha +\GI$).

The Labastida equation of motion enjoys the  Maxwell like gauge invariance
\be\nn
\delta\varphi=\DI^\alpha\xi_{\alpha}\, ,
\ee
thanks to the identity
\be\nn
(\DI_\alpha\DI^\alpha+ 
\GI \Big)\DI^\beta\xi_{\beta}=0\, ,
\ee
first uncovered by Baston~\cite{MR1165872}  . In fact the Labastida equation of motion has further gauge for gauge symmetries and accompanying
Bianchi for Bianchi identites. These are most easily displayed by writing the Labastida equation of motion in a form following from
the variation of an action. This is achieved by constructing the self-adjoint Einstein operator\footnote{The derivation of this result is described in~\cite{Cherney:2009mf,Campoleoni:2008jq} and 
amounts to composing the long operator with the Bessel series to balance its appearance on the right in~\eqref{BESSELR} 
and fixing $y$-independent representatives of 
${\rm coker} \, (y + {\bf g})$. }
\bea\nn
\  {\bf G}\ = \ \ :\!\frac{I_{1}(\sqrt{\GI\, \TR})}{2\,\sqrt{\GI\, \TR}}\!: \TR\, \Big(
\DI_\alpha\DI^\alpha{}+\GI\Big) \  
=\ \Big({\bm \delta}_\alpha{\bm \delta}^\alpha{}+\TR\Big)\,   \GI :\!\frac{I_{1}(\sqrt{\GI\, \TR})}{2 \, \sqrt{\GI\, \TR}}: \ \ = \ 
\bf{G}^* \, ,
\eea
in terms of which the Labastida equation of motion is equivalent to the ``Einstein'' equation of motion ${\bf G}\varphi=0$.

The Einstein operator has the compact, and manifestly self-adjoint expression
\be\nn
\shabox{$\begin{array}{rcl}
\bf{G} &=&\ : I_{0}(\sqrt{\GI\, \TR})\Big[\DI_\alpha\DE^\alpha + \DE_\alpha\DI^\alpha+2\Lambda\,({\bm N}-n)\Big]   \nonumber \\[6mm] 
&-&2\, \frac{I_{1}(\sqrt{\GI\, \TR})}{\sqrt{\GI\, \TR}}\Big[
(\DI_\alpha\DI^\alpha{}+\Lambda\, \GI) \, \TR + \GI\, ({\bm \delta}_\alpha{\bm \delta}^\alpha{}+\Lambda \, \TR )\Big]:\ \ \end{array}$
}
\ee
In all the above formul\ae, normal ordering denoted by $: \bullet :$
puts all factors of $\GI$ and $\TR$ to the far left and right, respectively
and
we have restored the dependence on the scalar curvature through  $\Lambda$ 
so that the $\Lambda \to 0$ hyperK\"ahler limit is manifest.
It is important to note that this operator acts on sections of $\wedge E$ of arbitrary degree. Therefore, the equation of motion we write down
is really the generating function for the equations valid at any degree and in arbitrary dimensions, this is what necessitates the operator-valued Bessel functions.

 Given the Einstein operator, we can now express the equations of motion, gauge and gauge for gauge invariances, Bianchi and Bianchi for Bianchi identities neatly in a single complex 
\be\label{BIG FORMAGGIO}
\begin{array}{c}
\cdots
\stackrel{\bm D} {\longrightarrow}
\wedge E \otimes \odot H
\stackrel {\bm D} {\longrightarrow}
\cdots 
\quad
\cdots
\stackrel{\bm F}{\longrightarrow} 
\wedge E \otimes \odot H
\stackrel{\bm F}{\longrightarrow} 
\cdots
\\
\ \Big|\hspace{-.8mm}\raisebox{-2.5mm}{\underline{\quad\quad\quad \ \raisebox{1mm}
{~~~~~~~~${\cal G}$~~~~}
\quad\qquad\quad }} \hspace{-1.2mm}{\Big\uparrow}
\end{array}
\ee
Here the operators ${\bf D}$ and ${\bf F}$ are closely related to the Dirac and Dirac--Fueter operators introduced by Baston~\cite{MR1165872} .
Explicitly, they act on sections of $\wedge E \otimes \odot H$ as
\bea
{\bf D}: \phi_{A_1\ldots A_k}{}^{\alpha_1\ldots \alpha_s} 
&\mapsto& s 
\nabla^{~\a}_{[A_1}\phi_{A_2\ldots A_{k+1}] \a}{}^{\alpha_1\ldots \alpha_{s-1}}\, ,\nn\\[3mm]
{\bf F}: \phi_{A_1\ldots A_k}{}^{\alpha_1\ldots \alpha_s}  
&\mapsto& 
k\nabla^{(\alpha_1}_A \phi^A{}_{ A_1\ldots A_{k-1}}{}^{\alpha_2\ldots \alpha_{s+1})}\, .
\eea
In an index free notation where 
$\Phi=\sum_{k,s}\phi^{\alpha_1\ldots \alpha_s}_{A_1\ldots A_k}\eta^{A_1}\cdots\eta^{A_k}\, z_{\alpha_1}\cdots z_{\alpha_s}$ $\in$ $\wedge E \otimes \odot H$,
we may simply write
\be\nn
{\bf D} = \eta^A \nabla_{\a A} \frac{\partial}{\partial z_\alpha}
={\bf d}_\a \frac{\pa}{\pa z_\a}\, ,\qquad
{\bf F} 
= z_\alpha \nabla^{\alpha  }_A \frac{\partial}{\partial \eta_A}
=z_\a {\bm \delta}^\a\, . 
\ee
Both these operators are nilpotent by virtue of the algebra~\eqref{qalg} and the identity $t^{\a\b}\psi_{\a\b \gamma_1\cdots \gamma_s}=0$.
Moreover, 
\be\nn
({\bf d}_\alpha {\bf d}^\alpha +{\bf g})\,  {\bf D} = 0 = {\bf F}\,  ({\bm \delta}_\alpha {\bm \delta}^\alpha+{\bf tr} )\, ,
\ee
verify the veracity of the complex~\eqref{BIG FORMAGGIO}. 

The incoming complex with differential ${\bf D}$ can be viewed as the quaternionic generalization of the Dolbeault complex~\cite{MR1165872},
while the outgoing complex with differential ${\bf F}$ is its dual ({\it i.e.} the Dirac--Fueter type operator ${\bf F}$ is a codifferential). Physically they encode
gauge invariances and Bianchi identities. The Einstein operator ${\cal G}$ gives the detour connecting the two complexes and, physically, the equations of
motion. Notice also, that it can connect the equations of motion at any degree in $\wedge E$ or $\odot H$, so gauge potentials are generic sections
of $\wedge E \otimes \odot H$. The mathematical elegance of this model is perhaps surprising, but even more remarkable is its {\it r\^ole} as the
arena for a minisuperspace quantization of ${\cal N}=2$ supersymmetric black holes. We further discuss this and other possible applications of our
theory in the conclusions.

\section{Conclusions}

The results presented in this paper rely on an  analogy between (i) differential forms on a K\"ahler manifold, (ii) tensors on a constant curvature manifold and (iii) the bundle
$$
\wedge E \otimes {\cal T} H 
$$
over a quaternionic manifold obtained by splitting its tangent bundle using the $\frak{sp}(2n)\otimes\frak{sp}(2)$ special holonomy and
then taking antisymmetric sections of the $\frak{sp}(2n)$ part~$E$ along with arbitrary $H$-tensors. The analogy with K\"ahler differential forms holds because the natural geometric operators on this bundle are in correspondence with the Dolbeault operators and the generators of the Lefschetz symmetry of Dolbeault cohomology. There is a relation to constant curvature manifolds  because, acting on sections of $\wedge E$, only the covariantly constant part
of the quaternionic K\"ahler Riemann tensor contributes. This means that the properties of the geometric operators we have studied are algebraically similar to the Lichnerowicz wave operator and the set of geometric operators that commute with it on a constant curvature manifold. In fact a main result of
this paper is the geometric calculus of operators, including a central wave operator, acting on $\Gamma(\wedge E \otimes {\cal T} H)$. Remarkably,
this seemingly purely mathematical structure was motivated by a study of supersymmetric black holes in four dimensional spacetime.

\begin{figure}
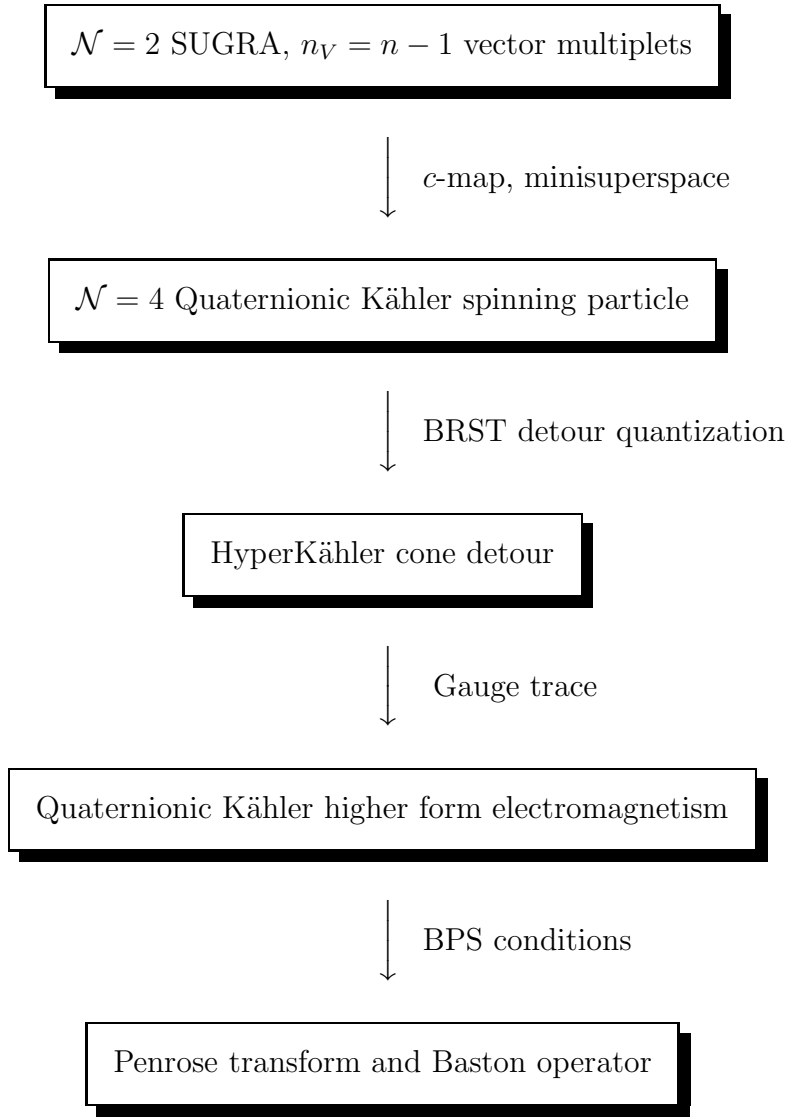

\begin{center}
\underline{\em \Large Physics Road Map}\\[8mm]

\begin{tabular}{cl}
\shabox{${\cal N}=2$ SUGRA, $n_V=n-1$ vector multiplets}\hspace{-3cm}\\[3mm]\\ 
\scalebox{1.3}{$\Big\downarrow$}\hspace{-3cm}& \hspace{-5cm}$c$-map, minisuperspace \\[3mm]\\
\shabox{${\cal N}=4$ Quaternionic K\"ahler spinning particle}\hspace{-3cm}\\[3mm]\\
\scalebox{1.3}{$\Big\downarrow$}\hspace{-3cm}& \hspace{-5cm}BRST detour quantization\\[3mm]\\
\shabox{HyperK\"ahler cone detour}\hspace{-3cm}\\[3mm]\\
\scalebox{1.3}{$\Big\downarrow$}& \hspace{-5cm} Gauge trace\\[3mm]\\
\shabox{Quaternionic K\"ahler higher form electromagnetism}\hspace{-3cm}\\[3mm]\\
\scalebox{1.3}{$\Big\downarrow$}& \hspace{-5cm}BPS conditions\\[3mm]\\
\shabox{Penrose transform and Baston operator}\hspace{-3cm}&
\end{tabular}
\end{center}
\caption{A map of the physical models encountered in this paper.\label{map}}
\end{figure}

The route from four dimensional black holes to a local quantum field theories on quaternionic K\"ahler manifolds is sketched in figure~\ref{map}.
It began with ${\cal N}=2$ SUGRA in four dimensions. Reducing along an isometry and specializing to spherical symmetry led to a spinning model
with four local worldline supersymmetries. Thanks to the $c$-map this spinning particle moves in a quaternionic K\"ahler manifold. Moreover, 
fermionic degrees of freedom  were retained in order that the BPS conditions of the spinning particle model corresponded to the reduced ones
of the four dimensional SUGRA, and therefore in turn to the linear evolution equations of the attractor mechanism. We then studied the 
quantization of this model through BRST detour methods. This led to the gauge invariant equation of motion~\eqref{LONG}. Let us make a few remarks on 
this model.

Given a $4n$-dimensional quaternionic K\"ahler manifold, it is always possible to find a $4n+4$ dimensional hyperK\"ahler manifold
whose metric is a quaternionic cone over the original $4n$-dimensional model~\cite{MR1096180,LeBrun,more}. In the work~\cite{MR1096180}, the dimensionally reduced supersymmetry parameters 
of the four dimensional SUGRA were shown to correspond to the extra four coordinates required to build a $4n+4$ dimensional 
hyperK\"ahler cone over the quaternionic K\"ahler, stationary, spherically symmetric, black hole moduli space. However, in BRST quantization the ghosts
correspond to the local gauge parameters, in particular the superghosts play the {\it r\^ole} of the supersymmetry parameters. Hence, the model~\eqref{LONG}, where we made no additional gaugings to eliminate ghosts, really should be viewed as a model on the hyperK\"ahler cone. This explains
the third signpost on the roadmap~\ref{map}. 

The next stop on the roadmap was motivated by ideas from higher spin models. In particular, our aim was to write down a model
where all ghosts had been eliminated from the physical cohomology. Based on ideas coming from our earlier work on orthosymplectic
constraint algebras, we suspected that gauging the Lefschetz--Verbitsky  trace operator would lead to a gauge invariant quantum field
theory generalizing both $p$-form electromagnetism and $(p,q)$-form K\"ahler electromagnetism to quaternionic K\"ahler manifolds.
This hunch was correct and led to the model~\eqref{BIG FORMAGGIO}. Interestingly enough, it could have been the case that this choice of route
would lead to a model that did {\it not} describe supersymmetric black holes. However, it is clear that in fact the quaternionic K\"ahler model
does so, and in a fascinating way. Examining the Labastida form of the equation of motion~\eqref{Labastida} we see that it is a product of the 
Baston operator and the Lefschetz--Verbitsky trace operator. As shown in~\cite{Neitzke:2007ke}, by explicity constructing the quaternionic Penrose transform
underlying Baston's quaternionic generalization of the Dolbeault complex, at least in the scalar sector of $\wedge E$,  zero modes of the Baston operator correspond to supersymmetric
black hole states. We suspect that within BRST quantization, this picture can be extended to a general correpsondence with the Baston complex. In this case,  solutions to our quaternionic K\"ahler electromagnetism theory would fall into two classes:
\begin{enumerate}
\item BPS solutions in the kernel of ${\bf d}_\alpha {\bf d}^\alpha + {\bf g}$.
\item Solutions whose non-vanishing image under ${\bf d}_\alpha {\bf d}^\alpha + {\bf g}$ lies in the kernel of ${\bf tr}$.
\end{enumerate}
This explains the last signpost of the roadmap~\eqref{map}. Clearly our work opens many avenues for further study:

Firstly, since our BRST quantization methods produce  a gauge theory on the hyperK\"ahler cone and furthermore rely on 
a polarization where one fourier transforms over half the ghost variables (alias quaternionic cone coordinates), there should 
exist a rather direct relationship between BRST quantization and the quaternionic twistor methods of~\cite{Neitzke:2007ke}.

Secondly, our quaternionic K\"ahler higher form electromagnetism may provide an interesting arena for further studies
of minisuperspace black hole quantization. One might hope that constructing interactions for this abelian gauge theory
could lead to a far more detailed understanding of these theories (perhaps along the lines of the multi-centered configuration and attractor flow trees---``third quantization''~\cite{Giddings:1988wv}).
This might sound extremely ambitious, since higher spin interactions are fraught with inconsistencies. 
However, it is possible
that some of the methods of Vasiliev, who has constructed three point higher spin interaction using a combination of unfolding techniques
(which are closely related to our BRST framework) and Chern--Simons like equations of motions based on a star product, could solve this problem. Also, we cannot help but remark, that whenever two seemingly disparate fields 
(such as higher spin interactions and four dimensional black hole physics) 
turn out to be related, oftentimes the flow of new ideas is bidirectional. In fact, we suspect that higher quantum corrections to ${\cal N}=2$ supergravities in four dimensions, could even have implications for possible higher spin interactions.

Finally, another topic that is worth further investigation is the novel Dirac operator in~\eqref{DIRAC}. This operator acts on the BRST superghost
Hilbert space; in the context of this paper it was merely a tool for constructing a nilpotent BRST charge. However, we suspect that 
it might have a distinguished {\it r\^ole}  to play. In particular, it would be fascinating to compute the Witten index of this operator.
Given that it was built from a supersymmetric quantum mechanical model, standard quantum methods may suffice for this.

\section*{Acknowledgements}

A.W. would like to thank Andy Neitzke and Boris Pioline for an early collaboration on this work, as well as many absolutely invaluable discussions. 
We would also like to thank Fiorenzo Bastianelli,  Roberto Bonezzi, Olindo Corradini, Dmitry Fuchs, Carlo Iazeolla and  
Albert Schwarz for useful discussions and comments.

\appendix

\end{document}